\documentclass[12pt,preprint]{aastex}

\newcommand{\hbeta} {H$\beta$}
\newcommand{\hgamma} {H$\gamma$}

\newcommand{\hepsilon} {H$\epsilon$}
\newcommand{\Te} {T_{\rm eff}}
\newcommand{\logg} {\log g} 
\begin{document}

\title{TOWARDS AN EMPIRICAL DETERMINATION OF THE ZZ CETI INSTABILITY STRIP}

\author{A. Gianninas, P. Bergeron, and G. Fontaine}
\affil{D\'epartement de Physique, Universit\'e de Montr\'eal, C.P.~6128, 
Succ.~Centre-Ville, Montr\'eal, Qu\'ebec, Canada, H3C 3J7.}
\email{gianninas@astro.umontreal.ca, bergeron@astro.umontreal.ca, 
fontaine@astro.umontreal.ca}

\begin{abstract}

We present atmospheric parameters for a large sample of DA white
dwarfs that are known to be photometrically constant. For each star,
we determine the effective temperature and surface gravity by
comparing high signal-to-noise ratio optical spectra to the predictions 
of detailed model atmosphere calculations. We also report the successful
prediction and detection of photometric variability in G232-38 based on
similar $\Te$ and $\logg$ determinations. The atmospheric parameters
derived for this sample of constant stars as well as those for the 
known sample of bright ZZ Ceti stars (now boosted to a total of 39)
have been obtained in a highly homogeneous way. We combine them to 
study the empirical red and blue edges as well as the purity of the ZZ
Ceti instability strip. We find that the red edge is rather well constrained
whereas there exists a rather large range of possibilities for the slope
of the blue edge.  Furthermore, the ZZ Ceti instability strip that
results from our analysis contains no nonvariable white dwarfs. Our
sample of constant stars is part of a much broader spectroscopic
survey of bright ($V < 17$) DA white dwarfs, which we have recently
undertaken. We also present here some preliminary results of this
survey. Finally, we revisit the analysis by Mukadam et al.~of the
variable and nonvariable DA stars uncovered as part of the Sloan
Digital Sky Survey. Their erroneous conclusion of an instability strip
containing several nonvariable stars is traced back to the low
signal-to-noise ratio spectroscopic observations used in that survey.

\end{abstract}


\section{INTRODUCTION}

The ZZ Ceti stars represent a class of variable white dwarfs whose
optical spectra are dominated by hydrogen lines (DA stars). They
occupy a narrow region in the $\Te$-$\logg$ plane known as the ZZ Ceti
instability strip, with an average effective temperature around
$\Te\sim11,600$~K and a width of roughly 1000~K. A precise
determination of the hot and cool boundaries of this instability strip
may eventually provide important constraints about the structure of
the outer layers of DA white dwarfs.  For instance, it has been
originally shown by \citet{winget82} that the location of the blue
edge is sensitive to the convective efficiency in the hydrogen zone,
which led \citet{fontaine84} to propose using this property as a
potential calibrator of the mixing-length theory in pulsating white
dwarfs. Similarly, the location of the red edge may help us understand
the mechanism responsible for the disappearance of the ZZ Ceti
phenomenon at low temperatures, which seems to be related to either
convective mixing of the hydrogen outer layer with the deep helium
envelope or the interaction of pulsation with convection
\citep{tassoul90}. Also of utmost importance is to determine whether all 
white dwarfs within the ZZ Ceti instability strip are pulsators. If the
strip is indeed pure, as first suggested by \citet{fontaine82}, ZZ Ceti
stars would necessarily represent a phase through which all DA stars
must evolve, and thus the results from asteroseismological studies might
provide constraints on the properties not only of known ZZ Ceti stars,
but on the whole population of DA stars as well. 

Determinations of the boundaries of the ZZ Ceti instability strip
prior to 1991 have been nicely summarized by \citet{wesemael91} who
discuss the results from various observational techniques, both
photometric and spectroscopic. Among the first photometric studies were
those conducted using Str\"omgren photometry by \citet{mcgraw79} and
later by \citet{fontaine85}. Both analyses made it evident that ZZ
Ceti stars formed a rather homogeneous class of DA white dwarfs in
color-color diagrams, a result that was not obvious from prior
analyses based on broad-band colors. Multichannel spectrophotometric
data of ZZ Ceti stars obtained by \citet{greenstein76} have been
analyzed by \citet{fontaine82},
\citet{greenstein82}, and by \citet{WK84} using slightly different absolute
flux calibrations.

Later on, \citet{wesemael86}, \citet{lamontagne87}, and
\citet{lamontagne89} have used ultraviolet spectra obtained by the IUE
satellite as an independent method of measuring the effective
temperature of ZZ Ceti stars. In their analysis they assumed a value of
$\logg=8$ for each star but also mentioned that this assumption could be a
source of uncertainty as several ZZ Ceti stars showed signs of having
$\logg$ significantly higher or lower (e.g., G226-29 and Ross 548,
respectively). Finally, \citet{daou90} have carried out the first
analysis of a set of ZZ Ceti stars using the spectroscopic technique
where optical spectroscopic observations of the individual Balmer lines
are fitted with synthetic spectra to obtain measures of both $\Te$ and
$\logg$. 

The effective temperatures for the ZZ Ceti stars inferred from these
photometric and spectroscopic studies are in fairly good agreement
according to Figure 1 of \citet{wesemael91}, with the blue edge in the
range $\Te=12,130-13,500$~K and the red edge in the range
$\Te=10,000-11,740$~K. However, this apparent agreement has been
seriously questioned by \citet{bergeron92b} who examined the effects of
different convective efficiencies on the optical spectra of DA white
dwarfs in the vicinity of the ZZ Ceti instability strip. The results
of their calculations showed that the predicted absolute fluxes, color
indices, and equivalent widths are sensitive to the convective efficiency 
in the range $\Te\sim 8000-15,000$~K, with a maximum
sensitivity around 13,000~K. Hence, without a detailed knowledge of
the convective efficiency in the atmosphere of ZZ Ceti stars, the
results from all previous photometric and spectroscopic analyses had
to be considered uncertain.

This problem of the convective efficiency in the atmosphere of ZZ Ceti
stars has been tackled by \citet[][B95 hereafter]{bergeron95c} who
used optical spectroscopic observations combined with UV energy
distributions to show that the so-called ML2/$\alpha=0.6$
parametrization of the mixing-length theory provides the best internal
consistency between optical and UV effective temperatures,
trigonometric parallaxes, $V$ magnitudes, and gravitational
redshifts. With the atmospheric convective efficiency properly
parameterized, the spectroscopic technique could now yield atmospheric
parameters $\Te$ and $\logg$ for the ZZ Ceti stars that were not only
accurate in a relative sense, but in an absolute sense as well. Hence
it was possible for the first time to demonstrate that the boundaries
of the ZZ Ceti instability strip were a function of both the effective
temperature and the surface gravity of the star. Our knowledge of the
boundaries of the ZZ Ceti instability strip prior to the study of
\citet{mukadam04b} discussed below is summarized in Figure 4 of
\citet{bergeron04}. The ZZ Ceti stars occupy a trapezoidal region in
the $\Te-\logg$ plane, with the blue edge showing a stronger
dependence on the surface gravity than the red edge does. Consequently
the width of the instability strip is also gravity-dependent, with
$\Delta\Te\sim800$~K at $\logg=7.5$ and nearly twice as wide at
$\logg=8.5$.

As mentioned above, the assessment of the purity of the instability
strip is also of considerable interest. More than twenty years ago,
\citet{fontaine82} have argued from their study of multichannel 
spectrophotometric data that the strip is most likely pure, and that
ZZ Ceti stars therefore represent an evolutionary phase through which
all DA white dwarfs must pass. This conclusion is strongly supported
by our spectroscopic analysis of the 36 known ZZ Ceti stars shown in
Figure 4 of \citet{bergeron04}. The latter also included 54
known nonvariable white dwarfs that were all found to lie clearly
outside the empirical instability strip. We note that, prior to this
effort, the purity of the instability strip had been questioned
repeatedly \citep{dolez91,kepler93,kepler95,silvotti97,giovannini98}.

More recently, \citet{mukadam04a} reported the discovery of 35 new ZZ
Ceti stars from the Sloan Digital Sky Survey (SDSS) along with a large
number of stars found to be photometrically constant. \citet{mukadam04b} 
used the results from this sample of both variable and nonvariable DA stars to
``redefine'' the location of the instability strip and to assess its
purity. Although their determinations of $\Te$ and
$\logg$ for these new ZZ Ceti stars place virtually all of the
variables within the instability strip defined by
\citet{bergeron04}, with a possible offset due to the use of a different
set of model spectra (see \S~3), they also found a large fraction of
nonvariable stars within the strip. These results are clearly at odds
with the conclusions from our work during the last ten years.

In this respect, we have been gathering over the recent years optical
spectroscopic observations for all known nonvariable DA white dwarfs
with the goal of (1) constraining the location of the boundaries of the
ZZ Ceti instability strip not only by analyzing the variable stars
within the strip itself, but also the photometrically constant stars in
its vicinity, and (2) increasing the statistical significance of the
purity of the empirical instability strip.  Some partial results from this
endeavor have been reported in \citet{bergeron04}. Here we present the
results of our entire sample in \S~2, which include the discovery of a
new ZZ Ceti star. In \S~3 we revisit the results of \citet{mukadam04b}
for the variable and nonvariable DA stars uncovered in the SDSS. We then
report in \S~4 on preliminary results of a much broader spectroscopy
survey of the white dwarf catalog of \citet{mccook99}. Our conclusions
follow in \S~5.

\section{PHOTOMETRIC SAMPLE}

\subsection{Spectroscopic Observations}

Our sample of photometrically constant DA stars is composed of 121
objects gathered from various sources. Firstly, we have searched the
literature for all mentions of DA white dwarfs observed in high-speed
photometry and where no variations were detected. These include two
Ph.D.~theses
\citep{mcgraw77,giovannini96} and several studies of the instability
strip including those of
\citet{dolez91},
\citet{kepler95}, and \citet{giovannini98}. Another source consists of 
previously unpublished data from various observing campaigns conducted
over the years by two of us (GF and PB) and collaborators. We were
also able to include 4 white dwarfs identified in the Hamburg Quasar
Survey and reported to be constant by \citet{mukadam04a}. Our sample
does not include, however, stars whose nonvariability has recently
come to our attention such as those reported by \citet{silvotti05} nor
those discovered in the SDSS by \citet{mukadam04a} and \citet{mullally05}.

Our sample of 121 nonvariable DA stars is listed in Table 1 in order
of increasing right ascension.  About 30\% of the spectra in this
sample were already available from the previous spectroscopic analyses
of \citet{BSL} and \citet{bergeron95a}. These spectra had been secured
using our standard setup at the Steward Observatory 2.3~m telescope
equipped with the Boller \& Chivens spectrograph. The $4.''5$ slit
together with the 600~line~mm$^{-1}$ grating blazed at 3568 \AA~in
first order provides a spectral coverage from about 3000 to 5250
\AA~at a resolution of $\sim$ 6 \AA~FWHM. An additional 40 spectra
were provided to us by C.~Moran (1999, private communication); these
have a comparable spectral coverage but at a slightly better
resolution of $\sim3$ \AA\ FWHM. Seven spectra from the southern
hemisphere are taken from the analyses of \citet{bragaglia95} and
\citet{BLR}. Finally, high signal-to-noise ratio (S/N) optical spectra for
36 objects were obtained specifically for the purpose of this project
during four observing runs in 2003 and 2004, using again the Steward
Observatory facility.

\subsection{Fitting Procedure}

The method used for fitting the spectroscopic observations relies on
the so-called spectroscopic technique developed by
\citet{BSL}, and which has been refined by B95 and more 
recently by \citet[][LBH hereafter]{liebert05}.  The most important
improvement of the method is the way the continuum used to normalize
individual Balmer lines is defined. The approach is slightly different
depending on the temperature range in question. For stars in the
interval 16,000 $\gtrsim \Te \gtrsim$ 9000 K, pseudo-Gaussian profiles
are used whereas outside this temperature range synthetic spectra are
utilized to determine the continuum (see Fig.~4 of LBH). Once
the Balmer lines are normalized properly, we proceed to fit them with
a grid of synthetic spectra derived from model atmospheres with a pure
hydrogen composition. Our grid covers a range between $\Te=1500$~K and
140,000~K by steps of 500 K at low temperatures ($\Te<17,000$~K) and
5000 K at high temperatures ($\Te>20,000$~K), and a range in $\logg$
between 6.5 and 9.5 by steps of 0.5 dex (steps of 0.25 dex are used
between 8000 K and 17,000 K where Balmer lines reach their maxima).
For models where convective energy transport becomes important, we
adopt the ML2/$\alpha=0.6$ parametrization of the mixing-length theory,
as prescribed by B95.

One of the trickiest aspects of fitting optical spectra near the ZZ
Ceti instability strip is the fact that we overlap the temperature
interval over which the equivalent widths of the Balmer lines reach
their maximum near $\Te\sim13-14,000$~K (see, e.g., Fig.~4
of B95). Hence, in some cases, the minimization procedure
allows two acceptable solutions, one on each side of this
maximum. When the true effective temperature of the star is more than
$\sim 2000$~K away from the maximum, it is possible from a simple
visual inspection of the fits to discriminate between the cool and the
hot solutions. Indeed, for identical equivalent widths, the Balmer
lines on the cool side of the maximum have deeper line cores. For
stars in the range $\Te\sim11,500-16,000$~K we rely on the slopes of
the observed and theoretical spectra normalized to unity at 4600~\AA\
to discriminate between both solutions. As the slope of the energy
distribution changes rapidly with temperature, it becomes relatively
easy to decide which solution to adopt. Finally, whenever possible,
our choice of solution has been confirmed by comparing multichannel,
Str\"omgren, or Johnson photometry published in
\citet{mccook99} with the theoretical color predictions of
\citet{bwb95}.

LBH used multiple spectroscopic observations of individual white
dwarfs to estimate the external uncertainties of the fitted
atmospheric parameters obtained from the spectroscopic technique (see
their Fig.~8). Their estimate of the external error of each fitted
parameter is 1.2\% in $\Te$ and 0.038 dex in $\logg$. We adopt the
same uncertainties in this analysis since both data sets are identical
in terms of data acquisition, reduction, and S/N.

\subsection{Results}

\subsubsection{Adopted Atmospheric Parameters}

The values of $\Te$ and $\logg$ for each of the 121 constant DA stars
are listed in Table 1. We also include masses and absolute visual
magnitudes derived from the evolutionary models of
\citet{wood95} with carbon-core compositions, helium layers of $q({\rm He})\equiv
M_{\rm He}/M_{\star}=10^{-2}$, and thick hydrogen layers of $q({\rm
H})=10^{-4}$. Several individual objects in Table 1 are worth
discussing before looking at the global properties of the sample.

There are four known unresolved double degenerates included in our
sample. The first three of those are G1-45
\citep[WD~0101+048;][]{maxted00}, and LP 550-52 
(WD~1022+050) and G21-15 \citep[WD~1824+050;][]{maxted99}. \citet{liebert91}
have shown that in such cases the atmospheric parameters derived are
in fact an average of the parameters of both components of the
system. Similarly, \citet{bergeron90a} suggested on the basis of
spectroscopic and energy distribution fits that GD 387 (WD~2003+437) is
probably composed of a DA and a DC star. They derived $\Te=14,340$ K
and $\logg=7.50$ for the DA component. Therefore, the atmospheric
parameters reported here for these four systems are quite uncertain.

Three stars in Table 1 have composite spectra, and reprocessing of the
EUV flux from the white dwarf primary in the chromosphere of the
secondary contaminates the center of some, or all, of the Balmer
lines.  These are PG 0308+096 \citep{saffer93}, PG 1643+144
\citep{kidder91} and Case 1 \citep[WD~1213+528; ][]{lanning82}. For PG
0308+096, the only contaminated line is
\hbeta. Therefore, we exclude that line from the fitting procedure
and are able to get a satisfactory fit with atmospheric parameters
identical to those reported in Table 2 of LBH.  Similarly, in the case
of PG 1643+144 we exclude both \hbeta~and \hgamma. For Case 1 however,
nearly all the spectral lines, and \hbeta\ in particular, are
contaminated by the companion. As before, we exclude \hbeta\ but we
also exclude 25 \AA\ from either side of the line centers for \hgamma\
through \hepsilon, fitting only the line wings of the Balmer
series. The effective temperature thus obtained, $\Te=13,920$~K agrees
well enough with that determined by \citet{sion84} based on a fit of
to the IUE spectrum, $\Te=13,000\pm500$ K,

Finally, our sample also includes two stars known to be magnetic, GD
77 \citep[WD~0637+477;][]{schmidt92} and G128-72
\citep[WD~2329+267;][]{moran98}. They both show the characteristic Zeeman splitting
of the Balmer lines caused by their magnetic fields and thus fitting
their spectra is problematic due to the additional spectral
broadening. Therefore the atmospheric parameters reported here for
these two objects remain uncertain. For instance we obtain for G128-72
a spectroscopic solution of $\Te=11,520$~K and $\logg=9.09$, while
a fit to the $BVRIJHK$ photometric energy distribution combined with a
trigonometric parallax measurement yields $\Te=9400$~K and
$\logg=8.02$ according to \citet{BLR}.

These uncertain atmospheric parameter measurements are indicated by
colons in Table 1 and we must pay particular attention to the
corresponding objects when discussing the ZZ Ceti instability strip
below.

\subsubsection{G226-29}

Before discussing the results of our analysis any further, we want to
consider the case of G226-29. Being the hottest ZZ Ceti star analyzed
by \citet{bergeron04} with $\Te=12,460$~K and $\logg=8.28$, G226-29
represents an important object for determining the slope of the blue
edge of the ZZ Ceti instability strip \citep[see Fig.~4
of][]{bergeron04}.  These atmospheric parameter determinations are
based on the same spectrum than the one used by B95 in their analysis
of the atmospheric convective efficiency in DA white dwarfs. However,
B95 also discuss a second spectroscopic observation of G226-29 with
derived atmospheric parameters that agree within the uncertainties
with the values given above. To be more specific, the atmospheric
parameters derived from this second observation are $\Te=12,260$~K and
$\logg=8.32$, consistent with the previous estimates within the
uncertainties quoted in the previous section. What is more interesting
perhaps is that this new temperature estimate is now in perfect
agreement with the UV temperature obtained from the IUE spectrum,
$\Te=12,270$~K (see Fig.~12 of B95). Given this improved internal
consistency, we adopt from now on these new atmospheric parameters for
G226-29.  These are reported in Table 2 together with the values for
the mass and absolute magnitude.

\subsubsection{New ZZ Ceti Stars}

To complete the picture, in addition to the nonvariable stars given in
Table 1, we need to include all ZZ Ceti stars for which we have
spectroscopic observations.  These include the 36 ZZ Ceti stars from
\citet{bergeron04} as well as 3 new ZZ Ceti stars: PB 520 and GD 133 discovered
by \citet{silvotti05} and Silvotti et al.~(2005, in preparation),
respectively, and G232-38 (WD~2148+539; $V=16.4$) discovered as part
of our ongoing spectroscopic survey of the McCook \& Sion catalog
described in
\S~\ref{sc:survey}. Our fits to the Balmer lines of these new
variables are presented in Figure \ref{fg:f1}; the atmospheric
parameters for each object are reported in Table 2 together with the
masses and absolute visual magnitudes. The values of $\Te$ and $\logg$
place PB 520 and G232-39 squarely within the limits of the ZZ Ceti
instability strip (see Fig.~\ref{fg:f6} below) and we were more
than confident that high speed photometric measurements would confirm
their variability.

\citet{silvotti05} had already reported the detection of photometric 
variability in PB 520. G232-38, on the other hand, had never been
observed before for photometric variability to our knowledge. Thus, we
obtained high-speed photometric observations of G232-38 during an
observing run in 2004 October at the 1.6 m telescope of the
Observatoire du mont M\'egantic equipped with LAPOUNE, the portable
Montr\'eal three-channel photometer. In all, we were able to obtain
3.9 h of data. Our sky-subtracted, extinction-corrected light curve of
G232-38 is displayed in Figure \ref{fg:f2}. G232-38 is clearly a ZZ
Ceti star with multiperiodic luminosity variations. The resulting
Fourier (amplitude) spectrum is displayed in Figure
\ref{fg:f3}. Three main-frequency components are easily discernible,
with periods of 741.6, 984.0 and 1147.4~s. These relatively long
periods and the rather large amplitude ($\lesssim 10$\%) of the
luminosity variations are consistent with the location of G232-38
somewhat closer to the red edge of the ZZ Ceti instability strip (see
below).

After this paper was submitted, it has come to our attention that GD
133 (WD~1116+026) has been been identified as a short period ($\sim
120$ s), low-amplitude ($< 1$\%) ZZ Ceti star by Silvotti et
al.~(2005, in preparation) based on high-speed photometric
observations obtained at the VLT with ULTRACAM. This object has long
been thought to be photometrically constant according to numerous
published sources \citep{mcgraw77,kepler95,giovannini96,silvotti97}.
Back in March 2003, two of us (G.F.~and P.B.) had even observed this
star with the 61-inch telescope at the Mount Bigelow observatory, the
light curve of which is displayed in Figure \ref{fg:f4}. Although
there is no obvious periodicity observed in the light curve, the
corresponding Fourier (amplitude) spectrum shown in Figure \ref{fg:f5}
yields one significant peak above the 1$\sigma$ noise level with a
period of 120.13~s, consistent with the observations of Silvotti et
al.  We have two independent optical spectra for GD 133, one from
C.~Moran (1999, private communication) that yields $\Te = 12,090$ K
and $\logg = 8.06$, and our own data obtained in 2003 June that yields
$\Te = 12,290$ K and $\logg = 8.05$. Although both sets of atmospheric
parameters are consistent within the uncertainties, the former
solution places GD 133 within the confines of our empirical
instability strip and this is the solution we will adopt here.  We
report the atmospheric parameters for GD 133 in Table 2 along with our
determination for the mass and absolute visual magnitude. We note that
the location of GD 133 at the blue edge of the strip (see Fig.~6
below) is entirely consistent with the low amplitude and short
pulsation period observed in Figures \ref{fg:f4} and \ref{fg:f5} .

\subsubsection{The Empirical ZZ Ceti Instability Strip}\label{sc:strip}

The locations of all 121 constant DA stars from Table 1 along with the
36 ZZ Ceti stars from \citet{bergeron04} and the 3 new ZZ Ceti stars
discussed above are plotted in Figure \ref{fg:f6} in a
$\Te-\logg$ diagram. Only 82 out the 121 nonvariables have atmospheric
parameters that place them within the confines of Figure
\ref{fg:f6}. The bold open circles within the strip correspond, 
from left to right, to the new ZZ Ceti stars GD 133, PB 520, and
G232-38.

Given this unbiased sample, we can clearly see that the ZZ Ceti stars
define a trapezoidal region in the $\Te$-$\logg$ plane in which no
nonvariable stars are found, within the measurement errors, in
agreement with the conclusions of \citet{bergeron04} and references
therein. And there is certainly no need here to go through any
statistical analysis to conclude that the ZZ Ceti instability strip is
indeed pure. We must also
note that all nonvariable white dwarfs claimed to be close or even
within the ZZ Ceti instability strip are in fact well outside the
strip according to our analysis.  These are GD 52
\citep[WD~0348+339;][]{dolez91,silvotti97}; G8-8 \citep[WD~0401+250;][]{silvotti97,kepler93};
GD 31 (WD~0231$-$054), Rubin 70 (WD~0339+523), GD 202 (WD~1636+160) according
to \citet{dolez91}; PB 6089 (WD~0037$-$006) and G130-5
(WD~2341$+$322) according to \citet{silvotti97}; BPM 20383 (WD~1053$-$550),
BPM 2819 (WD~0255$-$705) according to \citet[][PG 1022$+$050 is a double
degenerate]{kepler93}; PG 1119+385, GD 515 (WD~1654+637), GD 236
(WD~2226+061) according to \citet{kepler95}. There is also the case of GD
556 \citep[WD~2311+552;][]{dolez91,kepler95,giovannini98}, which we find
slightly hotter than the red edge of the strip; this object is
discussed further in the next section.

One of the primary goals of our study is to improve the determination
of the location of the blue and red edges of the empirical ZZ Ceti instability
strip by using both variable and nonvariable DA white dwarfs. The
results shown in Figure \ref{fg:f6} first reveal that the
location of the red edge is better constrained than the blue edge, in
particular because of the three nonvariables (GD 556, GD 426, and EC
12043-1337) that lie very close to the red edge. In contrast, there
are very few hot nonvariables near the blue edge. Note that the filled
squares at the top of the figure are unresolved double degenerates and
the atmospheric parameters obtained here are the average values of
both components of the system. Hence these cannot be used to constrain
the slope of the blue edge. In addition, our revised temperature for
G226-29, which is 200 K cooler than our previous estimate, now removes the
previous constraint we had on the slope of the blue edge. We show in
Figure \ref{fg:f6} the range of possibilities for the blue edge
as defined by our spectroscopic analysis. It is clear that additional
observations close to the blue edge are badly needed to constrain
the slope better.

We point out, in this connection, that nonadiabatic pulsation theory
does suggest that the slope of the blue edge in a $\Te-\logg$ diagram
such as the one shown in Figure \ref{fg:f6} should be significantly
smaller than that of the red edge, leading to an expected strip which
is wider at higher surface gravities. The last word on the question of
the theoretical ZZ Ceti instability strip has been presented by
\citet{fontaine03}. We show in Figure \ref{fg:f6} an updated comparison
with their theoretical results (solid lines). We find that the slope
of the theoretical blue edge is compatible with the range of
possibilities allowed by our empirical results. On the other hand, the
slope of the theoretical red edge is not too different from our own
determination however it is predicted to be somewhat hotter than the
red edge inferred from observation. Our aim in the future is to focus
on the $empirical$ boundaries with improved statistics, especially for
the blue edge.

A global characteristic that is also noticeable in Figure
\ref{fg:f6} is the trend toward higher values of $\logg$ as
$\Te$ decreases. This is now a familiar result observed in all
spectroscopic surveys extending to low temperatures (B95,
\citealt{SPY}, \citealt{kleinman04}, LBH, \citealt{gianninas05}). It has been
proposed by \citet{bergeron90b} that these high inferred masses could
be the result of small amounts of helium brought to the surface by the
hydrogen convection zone, hence increasing the atmospheric
pressure. When analyzed with pure hydrogen models, this increased
pressure could be misinterpreted as resulting from of a high mass
\citep[see also][]{boud05}.

\subsubsection{GD 556}

One constant star in Figure \ref{fg:f6}, GD 556, has an effective
temperature slightly hotter than the empirical red edge.
If we refer to Table 1, there are four
independent sources that concluded that GD 556 is not a variable DA
white dwarf. However, we would like to recall that initially G30-20
had also been found to be constant by
\citet{dolez91} and Bergeron \& McGraw (1989,
unpublished) but was later identified as a ZZ Ceti pulsator by
\citet{mukadam02}; GD 133 discussed above is also a good example. 
It is worth mentioning that GD 556 presents certain challenges as far
as photometric observations are concerned. Firstly, it is a rather dim
star with $V\sim16.2$
\citep{mccook99}. Secondly, its position near the red edge implies that 
if it is indeed a pulsator, it should show long period pulsations that
can be difficult to detect if one observes the star while two
pulsational modes are interfering destructively. Considering all
these facts, we believe that GD 556 is definitely worth re-observing
under favorable conditions, both photometrically and
spectroscopically. Nonetheless, if GD 556 truly is photometrically
constant, then considering our error bars, the fact that it lies
within the strip, albeit very close to the red edge, changes nothing
in our conclusions relative to the the purity of the ZZ Ceti
instability strip.

\section{RESULTS FROM THE SLOAN DIGITAL SKY SURVEY}

Since the discovery of the first pulsating DA white dwarf by
\citet{landolt68}, HL Tau 76, and up to the spectroscopic study of
\citet{bergeron04}, a total of 36 ZZ Ceti stars were known
\citep[see Table 1 of][]{bergeron04}, a quarter of which had been discovered
using the spectroscopic technique. In a single effort,
\citet{mukadam04a} reported the discovery of 35 {\it new} ZZ Ceti
pulsators, hence nearly doubling the number of known variables in this
class. Thirty three of these have been discovered in the white dwarf
SDSS sample, mostly from the first data release \citep{kleinman04},
and two more from the Hamburg Quasar Survey. Very recently,
\citet{mullally05} reported the discovery of eleven more ZZ Ceti stars
from SDSS as well as several nonvariable stars. However, the stars from
\citet{mullally05} are not included in the analysis and discussion that
follow.

ZZ Ceti candidates from the SDSS were selected for follow-up
high-speed photometry on the basis of various techniques including
$ugriz$ photometry, equivalent width measurements, and the
spectroscopic technique using SDSS spectra and Koester's model
atmospheres. By far, the spectroscopic technique led to a
significantly higher success rate of discovery than other techniques
(90\% by confining the candidates between $\Te=11,000$~K and
12,000~K). The 33 new SDSS pulsators are listed in Table 1 of
\citet{mukadam04a}, while nonvariables are given in 
their Tables 2 and 3 for different detection thresholds. An
examination of these tables reveal that all objects are relatively faint
($g\gtrsim 17$) due to the intrinsic characteristics of the Sloan
survey, which is aimed at identifying distant galaxies and
quasars. Stellar objects on a given plate with an assigned fiber had
to be faint in order not to saturate the detector.

Even though effective temperatures and surface gravities obtained from
spectroscopic fits were provided in their paper, \citet{mukadam04a}
did not discuss the implications of their new discoveries on the
empirical determination of the ZZ Ceti instability strip. That
discussion was deferred to a second paper by \citet{mukadam04b} who
analyzed in more detail the spectroscopic results from their first
paper, with a particular emphasis on the empirical ZZ Ceti instability
strip as inferred from the location of variables and nonvariables in
the $\Te-\logg$ plane. In particular, the authors of that study
question one more time the purity of the ZZ Ceti instability strip.
The results of their analysis are contrasted with our results in
Figure \ref{fg:f7}. We should mention that both analyses
rely on different sets of model atmospheres (ours versus D.~ Koester's
models) and there could be systematic offsets. But the most striking
feature of the Mukadam et al.~results is the large number of
nonvariable white dwarfs within their empirical instability strip.

Through a painstaking statistical analysis of their results,
\citet{mukadam04b} conclude that 18 nonvariables fall
within the ZZ Ceti instability strip. Given that 33 new pulsators have
been discovered from the same sample, the results suggest that the ZZ
Ceti instability strip is only $\sim 50$\% pure, at best. The authors
have even estimated the probability that the instability strip is pure
is only 0.004 \%! This result is of course in sharp contrast with our
conclusions based on a comparable number of white dwarfs, and
considerably brighter than those discovered in the SDSS. If indeed the
instability strip is contaminated by a significant fraction of
nonvariables, as implied by Mukadam et al., then the global properties
of DA stars inferred from asteroseismological studies of ZZ Ceti stars
could not be generalized to the entire population of
hydrogen-atmosphere white dwarfs as the ZZ Ceti pulsators would no
longer represent a phase through which {\it all} DA stars must
evolve. Another important implication of this challenging result is
that the pulsation instability of a white dwarf would no longer
depend solely on its effective temperature and stellar mass, but
would require an additional, yet unidentified, physical parameter to
discriminate variables and nonvariables within the instability strip.

How can our results be reconciled with those of Mukadam et al.?
The authors claim that since the discovery of white dwarf variables in
1968, their study represents the first analysis of a homogeneous set of
spectra acquired using the same instrument on the same telescope, and
with consistent data reductions. There is even an implicit suggestion
that this homogeneity could account for the fundamental difference
between their analysis and that of 
\citet{bergeron04}. However, this point of view completely ignores the
incentive behind the earlier study of B95 whose {\it
specific} goal was to provide an analysis of a homogeneous set of
spectroscopic observations of the 18 ZZ Ceti stars known at that time,
observable from the northern hemisphere. As discussed in \S~2.1 and
2.2 of B95, the first spectroscopic analysis of a sizeable
sample of ZZ Ceti stars by \citet{daou90} relied on spectra acquired
as part of a backup project by various observers, and thus with different
telescopes, spectrographs, detectors, and reduction procedures. As
such, the spectroscopic sample of Daou et al.~was somewhat inhomogeneous. To
overcome precisely this problem, it was deemed necessary for
B95 to reacquire optical spectra for the ZZ Ceti stars
using the same instrument setup and reduction techniques. Hence high
S/N spectroscopic observations for the 18 ZZ Ceti stars
were acquired using the 2.3-m telescope at Steward Observatory,
equipped with the Boller \& Chivens spectrograph and a Texas
Instrument CCD detector; spectra of four additional ZZ Ceti stars from
the southern hemisphere have also been analyzed by Bergeron et al.,
but even though the spectra were of comparable quality to those
obtained at the Steward Observatory, these four stars were treated
separately throughout their analysis to preserve the {\it homogeneity}
of the spectroscopic sample. 

Note that the same instrument setup has been used ever since in many
of our studies, and in particular in the recent extensive
spectroscopic analysis of LBH who reported effective temperatures and
surface gravities for nearly 350 DA stars drawn from the Palomar Green
(PG) Survey. Even though the completeness of the PG survey remains
questionable, the sample analyzed by Liebert et al.~represents one of
the largest statistically significant samples of DA stars analyzed to
date. Yet, only one ZZ Ceti candidate (PG 1349$+$552) was found within
the empirical instability strip together with 9 previously known ZZ
Ceti stars. High-speed photometric observations by \citet{bergeron04}
confirmed that PG 1349$+$552 was indeed a new ZZ Ceti pulsator. Hence
the conclusions of LBH are consistent with those of
\citet{bergeron04}, with the results presented in this paper, and with
the results of our ongoing survey of the McCook \& Sion catalog
discussed in \S~4, that {\it the empirical instability strip contains no
nonvariable stars}. Hence, arguments based on the homogeneity of the
spectroscopic analyses are unlikely to be able to explain the
discrepancy between our conclusions and the contrasting results of
\citet{mukadam04b}. If anything, a spectroscopic analysis of an
inhomogeneous data set should lead to a contamination of the
instability strip with nonvariables, not the other way around!

\citet{mukadam04b} also suggested that their analysis effectively samples 
a different population of stars, more distant by a factor of 10 than
that of the
\citet{bergeron04} sample. Actually, taking a median value of $g\sim18.5$ 
(see Fig.~\ref{fg:f8} below) and an absolute magnitude of $M_g=11.64$
obtained from a model atmosphere at $\Te=12,000$~K and $\logg=8$, we
derive a distance of only 230 pc, still relatively close by. There is
really no astrophysical reason to expect white dwarfs at that distance
to behave differently from those at shorter distances. Other
explanations must thus be sought.

A close examination of the 18 nonvariables claimed to be within the
instability strip by \citet[][see their Table 1]{mukadam04b} reveals
that all objects are among the faintest in their SDSS sample, as can
be seen from Figure \ref{fg:f8} where the distribution of SDSS
white dwarfs taken from Tables 1 to 3 of \citet{mukadam04a} is shown
as a function of the $g$ magnitude in the $ugriz$ photometric
system. As discussed by B95, the S/N of the spectroscopic
observations is one of the key aspects of the spectroscopic technique
for determining precise atmospheric parameters, the other important
one being the flux calibration. Since the exposure time of a given
SDSS spectrum is set by that of the entire plate, the corresponding
S/N must necessarily be a function of the magnitude of the star. To
verify this assertion, we have measured the S/N values\footnote{Here
the S/N is measured in the continuum between 4450 and 4750 \AA.} of
all SDSS spectra taken from
\citet{mukadam04a}, and plotted these values against the corresponding
$g$ magnitude. This is shown in Figure \ref{fg:f9}. As expected,
fainter stars have lower S/N spectra, and only objects brighter than
$g\sim17$ have S/N above 40.  This is not the case with standard slit
spectroscopy, however, where the exposure time can be adjusted on a
star-to-star basis. In B95 for instance, the exposure times were set to
achieve an imposed lower limit of ${\rm S/N}\sim80$, although most
spectra had ${\rm S/N}\gtrsim 100$ since the exposure times were also
set long enough to cover several pulsation cycles (for an average of
$\sim4.8$ cycles) in order to obtain meaningful time-averaged spectra.
We mention that this last criterion is not necessarily met in the SDSS
spectroscopic data.

We now turn to a more detailed comparison of S/N between our
spectroscopic sample and that of \citet{mukadam04a}. We show in the
top panel of Figure \ref{fg:f10} the distribution of S/N values
for our spectroscopic sample, including the photometrically constant
stars from Table 1, the 36 ZZ Ceti stars from
\citet{bergeron04}, and the 3 new pulsators from Table 2.  Out of 39 ZZ
Ceti stars, 12 have spectra with an admittedly lower S/N value than
the imposed lower limit of $\sim80$ set by B95 in their
analysis. These spectra correspond to data provided to us by C.~Moran
(1999, private communication) in the course of his search for double
degenerate binaries, or to spectroscopic observations obtained prior
to the discovery of the photometric variability of the object; this
includes the two ZZ Ceti stars PB 520 and G232-38 analyzed in this
paper. Still, only two ZZ Ceti stars have spectra with ${\rm S/N}<50$,
and none below 40. The spectra for our photometrically constant sample
also have fairly high S/N values, almost all above 50. In contrast,
the S/N of the SDSS spectra\footnote{Only a fraction of the SDSS
spectra could be recovered from the SDSS Web site.} shown in the
bottom panel of Figure
\ref{fg:f10} have considerably lower values, with most spectra having
${\rm S/N}<60$. Even worse, the subsample of nonvariables claimed to
lie within the instability strip (hatched histogram) has even lower
S/N values, with most objects having ${\rm S/N}<40$. Hence, it is
perhaps not too surprising that the results of \citet{mukadam04b}
regarding the purity of the ZZ Ceti instability strip, which are based
on an analysis of low S/N spectra, differ so much from our own
conclusions based on much higher quality spectroscopic observations.

We illustrate in Figure \ref{fg:f11} a typical spectrum from our own
sample of ZZ Ceti stars, GD 66 with ${\rm S/N}=80$, and one from the
SDSS sample, SDSS J084746.81$+$451006.3 with ${\rm S/N}=20$. The S/N
value of the latter is actually more typical of the sample of
nonvariables found within the strip (bottom panel of
Fig.~\ref{fg:f10}). It is clear that the spectroscopic solution will
necessarily depend on the quality of these spectra.  To quantify this
assertion, we performed a Monte Carlo simulation by taking a series of
model spectra at $\Te=12,000$ K and $\logg=8.0$, by adding random
noise to achieve a given signal-to-noise ratio, and by fitting these
spectra with our standard fitting procedure. The resulting $\Te$ and
$\logg$ values are then used to compute the standard deviations
$\sigma_{\Te}$ and $\sigma_{\logg}$ for this assumed S/N value.
Values of S/N from 10 to 200 were explored, thus encompassing the
entire range exhibited by the spectra analyzed in this paper and by
\citet{mukadam04b}. The results of this exercise are displayed in
Figure \ref{fg:f12}. It is clear that stars with low S/N spectra will
yield atmospheric parameters with larger internal uncertainties than
those derived from higher quality observations. In particular, if we
again take S/N = 20 as indicative of the SDSS stars, we see from
Figure \ref{fg:f12} that such spectra would yield effective
temperatures that are uncertain by $\sim 500$ K. If we consider that
the width of the empirical instability strip is $\sim 1000$ K, it is
easy to understand how lower quality spectra could easily place
non-variable stars within the strip and vice-versa. Furthermore, stars
with S/N $\sim 80$, typical of our photometric sample, exhibit
uncertainties of roughly 150 K, which is entirely consistent with the
uncertainties quoted in \S 2.2. Thus despite the homogeneous
characteristics of the SDSS spectra in terms of instrument, telescope,
and data reductions, their typical S/N is most likely too low to allow
a precise measurement of the atmospheric parameters for these stars,
or to determine accurately the location of the empirical ZZ Ceti
instability strip, or to assess the purity of the strip for that
matter.

Finally, we examine in Figure \ref{fg:f13} the location of the SDSS
white dwarfs in the $\Te-\logg$ plane. In each panel we consider only
the objects with spectra above a certain threshold in S/N
(the bottom panel includes all objects). Also reproduced is the
empirical ZZ Ceti instability strip determined by \citet{bergeron04}.
The top panel with ${\rm S/N}>70$ corresponds to a threshold that
would include $\sim80$\% of all ZZ Ceti stars from the sample of
\citet[][top panel of Fig.~\ref{fg:f10}]{bergeron04}. 
By comparison, only 2 objects from the SDSS sample meet this
criterion.  Hence if we restrict the analysis to the best spectra of
both samples, the results are consistent: all variables are found
within the empirical strip and all nonvariables lie outside. For ${\rm
S/N}>40$ (middle panel), 9 objects from the SDSS sample are found in
the temperature range shown in Figure \ref{fg:f13}. In this case,
however, one variable star (WD~1711+6541 at $\Te=11,310$~K and
$\logg=8.64$) falls slightly below the empirical red edge of the
strip, while one nonvariable star (WD~1338$-$0023 at $\Te=11,650$~K
and $\logg=8.08$) sits comfortably near the middle of the strip.
Since these two objects are relatively bright ($g=16.89$ and 17.09,
respectively), we managed to secure our own spectroscopic observations
of these stars using the Steward Observatory 2.3 m telescope during an
observing run in 2004 May. Reassuringly enough, our independent
analysis of these two objects in terms of both data and models ---
$\Te=11,490$~K and $\logg=8.56$ for WD~1711+6541 and $\Te=11,980$~K
and $\logg=7.94$ for WD~1338$-$0023 --- places them where they are
expected, that is, inside and outside the instability strip,
respectively (see Fig.~\ref{fg:f13}).

For completeness, we show at the bottom of Figure \ref{fg:f13} all the
objects from the SDSS sample (${\rm S/N}>0$). Once again, we can see
that the bulk of this sample is characterized with S/N values below
40, the threshold value used in the middle panel, and that the
conclusion about the purity of the ZZ Ceti instability strip rests
heavily on the quality of the spectroscopic observations.

\section{ONGOING SPECTROSCOPIC SURVEY}\label{sc:survey}

In order to increase the number of stars in our spectroscopic (and
eventually photometric) sample, we have undertaken a broader
spectroscopic survey of DA stars drawn from the Catalog of
Spectroscopically Identified White Dwarfs of \citet{mccook99}. We have
defined our sample using the following criteria: (1) a temperature
index lying between 3 and 7, (2) apparent visual magnitudes of V $<$
17 and (3) declination greater than $-$30 degrees.  High S/N optical
spectroscopic observations are currently being secured for each star
that meets these criteria. This survey was initiated with several
goals in mind. First and foremost, we wish to obtain measurements of
$\Te$ and $\logg$ for each object. Secondly, we want to confirm the
spectroscopic classification of stars from the catalog (we have
already identified 29 stars misclassified as DA stars that are clearly
lower gravity objects). A final goal of our survey is to identify new
ZZ Ceti candidates (G232-38 has been discovered in this survey). This
is the reason for restricting ourselves to stars with the
aforementioned range of temperature indices.  Some preliminary results
of this analysis have already been presented in
\citet{gianninas05}. Among these is the discovery of a unique DAZ
white dwarf, GD 362 \citep{gianninas04}.

The combined results of our ongoing spectroscopic survey and of
the photometric sample analyzed in \S~2 are displayed in Figure
\ref{fg:f14} as triangles and circles, respectively. The three 
low-gravity objects in the vicinity of the instability strip are known
double degenerate systems. We have already discussed two of these (see
\S~\ref{sc:strip} and Fig. \ref{fg:f6}), the third is GD 429 
\citep{maxted00} which has yet to be observed for photometric variability. 
We clearly see that many objects from our survey lie very close to
both the red and blue edges of the instability strip. These stars are
important as we attempt to determine better the exact boundaries of
the instability strip. Therefore, we plan on securing high speed
photometric observations for these objects in order to confirm their
photometric status. These results will be reported in due time.

\section{CONCLUSION}

We have gathered optical spectra for 121 photometrically constant DA
white dwarfs for which we derived values of $\Te$ and $\logg$. Using
these nonvariable white dwarfs together with a sample of 39
relatively bright ZZ Ceti stars, we wished to obtain a better
understanding of the location and shape of the red and blue edges of
the ZZ Ceti instability strip. In so doing, we have succeeded in
better populating the $\Te$-$\logg$ plane in the vicinity of the ZZ
Ceti instability strip. We find that the location and slope of the red
edge is quite well constrained whereas our newly adopted atmospheric
parameters for G226-29 allow for a much broader range of slopes for
the blue edge which would accommodate our current photometric
sample. Furthermore, we find no nonvariable white dwarfs within the ZZ
Ceti instability. This supports our belief that ZZ Ceti stars
represent an evolutionary stage by which all DA white dwarfs must
pass.

The optical spectra that we analyzed were gathered as part of a more
extensive survey of DA white dwarfs from the catalog of
\citet{mccook99}.  This survey has several goals, among them, the
identification of candidate ZZ Ceti stars. Thus far, two of these, PB
520 and G232-38, have been identified as ZZ Ceti pulsators by
\citet{silvotti05} and us, respectively. The spectroscopic
technique pioneered by B95 has proven once again to be an invaluable
tool as far as identifying new candidate ZZ Ceti stars. Indeed, it has
maintained its 100\% success rate in predicting variability in DA
white dwarfs. With the inclusion of PB 520, G232-38, and GD 133 the
number of ZZ Ceti stars (excluding the SDSS stars) swells to 39 of
which 11 have been successfully identified using this method. Even
among the ZZ Ceti stars discovered through SDSS, the most fruitful
method for identifying candidates was the spectroscopic method
\citep{mukadam04a}. In the future, in order to define better the blue
edge, and to study further the instability strip as a whole, we plan on
securing high speed photometric observations for all the DA white dwarfs
that are within or near the boundaries of the empirical strip and that have
never been observed for variability.

We have also been able to show the importance of using high-quality
data (i.e., high S/N) when performing analyses such as these through an
in-depth examination of the data used by \citet{mukadam04b} in their
study. It is clear that their controversial results, which place a large
number of nonvariable stars within the instability strip, can be traced
back to spectra of lesser quality that greatly affect the result of the
spectroscopic fit. However, one cannot discount the fact that
\citet{mukadam04a} have nearly doubled the number of known ZZ Ceti
stars as well as adding a large number of nonvariable DA white dwarfs
to the mix. The study of the ZZ Ceti instability strip can only
benefit from the inclusion of all recently identified variable and
nonvariable DA white dwarfs within our sample. We are therefore
exploring the possibility of obtaining high S/N optical spectra for
all the DA white dwarfs from \citet{mukadam04a}, \citet{silvotti05}, 
and \citet{mullally05} in the near future.

We would like to thank the director and staff of Steward Observatory
for the use of their facilities. We would also like to thank the
director and staff of the Observatoire du mont M\'egantic for the use of
their facilities and for supporting LAPOUNE as a visitor instrument.
We also acknowledge the contribution of F.~Provencher in the analysis
of the SDSS data. This work was supported in part by the NSERC Canada
and by the Fonds Qu\'eb\'ecois de la recherche sur la nature et les
technologies (Qu\'ebec). GF acknowledges the contribution of the Canada
Research Chair Program.

\clearpage
\clearpage
\begin{deluxetable}{clrcccc}
\tabletypesize{\scriptsize}
\tablecolumns{7}
\tablewidth{0pt}
\tablecaption{Atmospheric Parameters of Photometrically Constant DA White Dwarfs}
\tablehead{
\colhead{WD} &
\colhead{Name} &
\colhead{$T_{\rm eff}$ (K)} &
\colhead{$\log g$} &
\colhead{$M/M_{\odot}$} &
\colhead{$M_V$} &
\colhead{Sources}}
\startdata
0005$-$163    &G158-132          &14160&7.79     &0.50     &11.03    &1                \\
0009+501      &G217-37           & 6610&8.36     &0.83     &14.31    &2                \\
0011+000      &G31-35            & 9640&8.16     &0.70     &12.53    &3                \\
0030+444      &G172-4            &10370&8.20     &0.73     &12.33    &3                \\
0032$-$175    &G226-135          & 9830&8.18     &0.71     &12.49    &4                \\
\\
0033+016      &G1-7              &10980&8.83     &1.12     &13.27    &5, 6             \\
0037$-$006    &PB 6089           &14920&7.86     &0.54     &11.04    &1, 2, 5, 7, 8    \\
0101+048\tablenotemark{a}      &G1-45             & 8530:&8.27:     &0.77     &13.17    &4                \\
0103$-$278    &G269-93           &13290&7.83     &0.52     &11.21    &6, 9             \\
0115+521      &GD 275            &10710&8.12     &0.68     &12.09    &1                \\
\\
0135$-$052    &L870-2            & 7280&7.85     &0.51     &13.18    &4                \\
0143+216      &G94-9             & 9290&8.49     &0.92     &13.22    &3                \\
0148+467      &GD 279            &13430&7.93     &0.57     &11.33    &3                \\
0151+017      &G71-41            &12330&7.89     &0.54     &11.42    &1, 6, 8          \\
0208+396      &G74-7             & 7340&8.10     &0.66     &13.49    &2                \\
\\
0213+396      &GD 25             & 9320&8.56     &0.96     &13.32    &3                \\
0231$-$054    &GD 31             &13550&8.66     &1.02     &12.46    &3, 5             \\
0238+333      &KUV 02386+3322    &13390&8.23     &0.75     &11.77    &10               \\
0243+155      &PG 0243+155       &16670&8.02     &0.63     &11.08    &5                \\
0255$-$705    &BPM 02819         &10560&8.10     &0.66     &12.11    &3, 6, 11         \\
\\
0302+621      &GD 426            &11000&8.21     &0.73     &12.15    &5                \\
0308+096\tablenotemark{b}      &PG 0308+096       &25900&8.08     &0.68     &10.36    &9                \\
0326$-$273    &LTT 1648          & 9250&7.86     &0.52     &12.24    &2, 9             \\
0332+320      &G38-4             &10370&8.13     &0.68     &12.21    &3                \\
0339+523      &Rubin 70          &12640&7.39     &0.33     &10.69    &5, 6, 8, 9       \\
\\
0339$-$035    &GD 47             &12470&7.98     &0.60     &11.53    &3, 5             \\
0348+339      &GD 52             &14190&8.20     &0.74     &11.63    &1, 7             \\
0352+096      &HZ 4              &14030&8.19     &0.73     &11.64    &11               \\
0401+250      &G8-8              &12240&7.99     &0.60     &11.57    &1, 5, 6, 7, 8, 11\\
0406+169      &LB 227            &15070&8.26     &0.78     &11.62    &12               \\
\\
0407+179      &HZ 10             &13620&7.79     &0.50     &11.11    &3                \\
0418+153      &LB 212            &13480&7.99     &0.61     &11.41    &5                \\
0440+510      &G175-46           & 8620&8.22     &0.74     &13.04    &3                \\
0453+418      &GD 64             &13660&7.68     &0.44     &10.94    &3                \\
0513+756      &GD 433            &13540&7.76     &0.48     &11.08    &1, 5             \\
\\
\\
\\
\\
\\
0518+005      &GD 67             &13340&7.88     &0.55     &11.28    &3                \\
0533+322      &G98-18            &10680&7.89     &0.54     &11.76    &1, 5             \\
0637+477\tablenotemark{c}      &GD 77             &14000:&8.21:     &0.74     &11.66    &2                \\
0710+216      &GD 83             &10480&8.07     &0.65     &12.09    &3                \\
0743+442      &GD 89             &14500&8.36     &0.84     &11.85    &1                \\
\\
0816+387      &G111-71           & 7700&8.07     &0.64     &13.26    &4                \\
0830+371      &G115-9            & 9180&8.26     &0.76     &12.87    &3                \\
0839$-$327    &LHS 253           & 9270&7.89     &0.54     &12.27    &3                \\
0913+442      &G116-16           & 8680&8.20     &0.72     &12.98    &3                \\
0920+216      &LB 3025           &18000&7.83     &0.53     &10.66    &5                \\
\\
0926$-$039    &G161-36           &12860&7.86     &0.53     &11.32    &8                \\
0928$-$713    &BPM 05639         & 8580&8.28     &0.78     &13.16    &3                \\
0943+441      &G116-52           &12820&7.55     &0.39     &10.90    &6, 8, 11         \\
0950+077      &PG 0950+078       &14770&7.95     &0.59     &11.19    &12               \\
0950$-$572    &BPM 19738         &12400&7.68     &0.44     &11.13    &3, 8             \\
\\
0955+247      &G49-33            & 8620&8.30     &0.79     &13.18    &3                \\
0956+045      &PG 0956+046       &18150&7.81     &0.52     &10.62    &5                \\
1022+050\tablenotemark{a}      &LP 550-52         &11680:&7.64:     &0.42     &11.20    &6, 8, 11         \\
1026+023      &LP 550-292        &12500&7.95     &0.58     &11.49    &1, 6, 8, 9       \\
1046+281      &Ton 547           &12610&7.97     &0.59     &11.51    &3, 7             \\
\\
1053$-$550    &BPM 20383         &13420&7.81     &0.51     &11.16    &11               \\
1101+364      &PG 1101+364       &13040&7.24     &0.29     &10.41    &6, 9             \\
1108+475      &GD 129            &12460&8.24     &0.76     &11.92    &13               \\
1119+385      &PG 1119+386       &16500&7.94     &0.58     &10.98    &6                \\
1122+546      &GD 307            &14380&7.83     &0.52     &11.06    &1                \\
\\
1147+255      &G121-22           &10200&8.14     &0.69     &12.29    &3, 6             \\
1204$-$136    &EC 12043$-$1337   &11180&8.24     &0.76     &12.16    &13               \\
1213+528\tablenotemark{b}      &Case 1            &13920&8.16     &0.71     &11.60    &11               \\
1229$-$012    &HE 1229$-$0115    &19740&7.52     &0.41     &10.05    &5, 7             \\
1241+235      &LB 16             &26730&7.93     &0.60     &10.05    &7                \\
\\
1244+149      &G61-17            &10680&8.06     &0.64     &12.02    &6                \\
1253+482      &GD 320            &13970&7.59     &0.41     &10.78    &1, 8             \\
1327$-$083    &Wolf 485A         &13920&7.86     &0.54     &11.17    &1, 6, 8, 11, 14  \\
1418$-$005    &PG 1418$-$005     &14290&7.82     &0.51     &11.06    &6, 8             \\
1431+153      &PG 1431+154       &13550&7.95     &0.58     &11.35    &12               \\
\\
\\
\\
\\
\\
1448+077      &G66-32            &14170&7.75     &0.48     &10.97    &6, 8             \\
1507$-$105    &GD 176            &10100&7.75     &0.47     &11.76    &3, 6             \\
1508+637      &GD 340            &10450&8.12     &0.68     &12.18    &1                \\
1510+566      &G201-39           & 9240&8.13     &0.68     &12.63    &3                \\
1531+184      &GD 186            &13220&7.89     &0.55     &11.30    &3                \\
\\
1537+651      &GD 348            & 9740&8.15     &0.69     &12.47    &3, 7             \\
1539$-$035    &GD 189            &10080&8.30     &0.79     &12.59    &3, 6             \\
1544$-$377    &L481-60           &10580&8.09     &0.66     &12.09    &8                \\
1550+183      &GD 194            &14260&8.25     &0.77     &11.70    &6                \\
1555$-$089    &G152-B4B          &13960&7.83     &0.52     &11.12    &6, 8             \\
\\
1606+422      &Case 2            &12690&7.74     &0.47     &11.17    &2, 6, 8          \\
1609+135      &G138-8            & 9320&8.64     &1.01     &13.48    &3                \\
1636+160      &GD 202            &13620&7.81     &0.51     &11.13    &5, 6             \\
1637+335      &G180+65           &10150&8.17     &0.71     &12.35    &3                \\
1643+143\tablenotemark{b}      &PG 1643+144       &26850&7.91     &0.60     &10.02    &12               \\
\\
1654+637      &GD 515            &15070&7.63     &0.43     &10.70    &6                \\
1655+215      &G169-34           & 9310&8.20     &0.73     &12.72    &2, 3             \\
1706+332      &G181-B5B          &12960&7.80     &0.50     &11.21    &3                \\
1716+020      &G19-20            &13210&7.77     &0.49     &11.13    &3, 6, 8          \\
1743$-$132    &G154-85B          &12300&7.88     &0.54     &11.42    &3                \\
\\
1824+040\tablenotemark{a}      &G21-15            &11970:&7.57:     &0.39     &11.06    &5                \\
1826$-$045    &G21-16            & 9210&8.16     &0.70     &12.70    &3                \\
1827$-$106    &G155-19           &13300&7.63     &0.42     &10.93    &8                \\
1840$-$111    &G155-34           &10170&8.23     &0.75     &12.44    &3, 6             \\
1857+119      &G141-54           & 9920&8.12     &0.68     &12.36    &3, 6             \\
\\
1911+135      &G142-B2A          &13270&7.85     &0.53     &11.25    &6, 9             \\
1952$-$206    &LTT 7873          &13740&7.85     &0.53     &11.18    &2, 3             \\
1953$-$011    &G92-40            & 7780&8.25     &0.75     &13.49    &3                \\
2003+437\tablenotemark{a}      &GD 387            &16910:&7.80:     &0.51     &10.73    &1                \\
2025+488      &GD 390            &10720&8.05     &0.63     &11.99    &1                \\
\\
2029+183      &GD 230            &13090&7.79     &0.49     &11.18    &3                \\
2047+372      &G210-36           &14070&8.21     &0.74     &11.66    &8                \\
2059+190      &G145-4            & 6980&8.42     &0.86     &14.18    &2                \\
2105$-$820    &BPM 01266         &10620&8.25     &0.76     &12.33    &3, 6, 8, 11      \\
2115$-$560    &BPM 27273         & 9760&8.13     &0.68     &12.43    &6                \\
\\
\\
\\
\\
\\
2117+539      &G231-40           &13990&7.78     &0.49     &11.04    &9                \\
2124+550      &G231-43           &13340&8.34     &0.82     &11.95    &3, 9             \\
2126+734      &G261-43           &15290&7.84     &0.53     &10.97    &6, 11            \\
2136+229      &G126-18           &10210&8.10     &0.67     &12.23    &6                \\
2149+372      &GD 397            &13080&7.87     &0.54     &11.29    &1, 6             \\
\\
2226+061      &GD 236            &15280&7.62     &0.43     &10.66    &6                \\
2246+223      &G127-58           &10650&8.80     &1.10     &13.32    &3, 6             \\
2258+406      &G216-B14B         & 9860&8.23     &0.75     &12.55    &6                \\
2306+130      &KUV 23060+1303    &13250&7.92     &0.56     &11.34    &9                \\
2311+552      &GD 556            &11180&8.15     &0.69     &12.01    &1, 5, 6, 8       \\
\\
2314+064      &PB 5312           &17570&7.98     &0.61     &10.93    &1, 5             \\
2322+206      &PG 2322+207       &13060&7.84     &0.52     &11.26    &2, 9             \\
2329+267\tablenotemark{c}      &G128-72           &11520:&9.09:     &1.24     &13.67    &2, 3             \\
2337$-$760    &BPM 15727         &13420&7.39     &0.33     &10.57    &6                \\
2341+322      &G130-5            &12570&7.93     &0.57     &11.45    &1, 3, 5, 6, 7, 8 \\
\\
2351$-$335    &LDS 826A          & 8850&8.27     &0.77     &13.02    &2                \\
\enddata
\tablenotetext{a}{Double degenerate}
\tablenotetext{b}{Composite spectrum}
\tablenotetext{c}{Magnetic}
\tablecomments{
(1) G. Fontaine (1979-1984, unpublished); 
(2) P. Bergeron \& J. T. McGraw (1989, unpublished); 
(3) \citet{mcgraw77}; 
(4) \citet{kanaan02}; 
(5) G. Vauclair (1979-1999, unpublished), \citet{dolez91}; 
(6) \citet{kepler95}; 
(7) \citet{silvotti97}; 
(8) \citet{giovannini96}; 
(9) P. Bergeron \& G. Fontaine (1990, unpublished); 
(10) G. Fontaine \& P. Bergeron (1999, unpublished); 
(11) \citet{kepler93}; 
(12) \citet{mukadam04a}; 
(13) G. Fontaine \& P. Bergeron (2003, unpublished); 
(14) \citet{wesemael85}. }
\end{deluxetable}

\clearpage
\clearpage
\begin{deluxetable}{llcccc}
\tabletypesize{\footnotesize}
\tablecolumns{6}
\tablewidth{0pt}
\tablecaption{Atmospheric Parameters of ZZ Ceti Stars}
\tablehead{
\colhead{WD} &
\colhead{Name} &
\colhead{$T_{\rm eff}$ (K)} &
\colhead{$\log g$} &
\colhead{$M/M_{\odot}$} &
\colhead{$M_V$}}
\startdata
1039+412      &PB 520            &11550   &8.10     &0.66     &11.85                  \\
1116+026      &GD 133            &12090   &8.06     &0.64     &11.70                  \\
1647+591\tablenotemark{a}      &G226-29           &12260   &8.31     &0.80     &12.10                  \\
2148+539      &G232-38           &11350   &8.01     &0.61     &11.76                  \\
\enddata
\tablenotetext{a}{Based on a new spectroscopic observation (see text).}
\end{deluxetable}

\clearpage

\figcaption[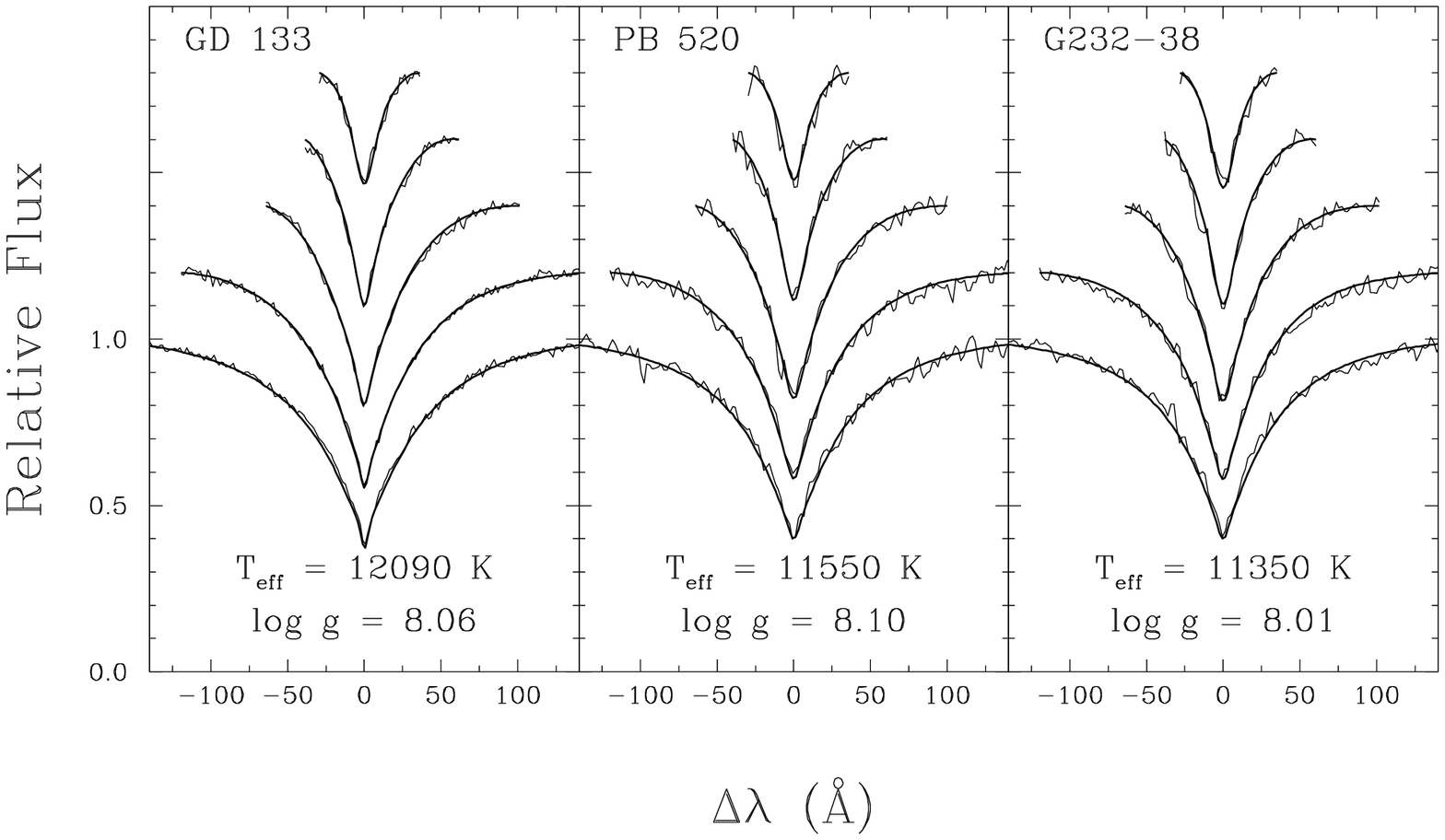] 
{Model fits to the individual Balmer line profiles of GD 133, PB 520, and
G232-38. The lines range from \hbeta~({\it bottom}) to H8 ({\it top}),
each offset vertically by a factor of 0.2. Values of $\Te$ and $\logg$
have been determined from ML2/$\alpha=0.6$ models. \label{fg:f1}}

\figcaption[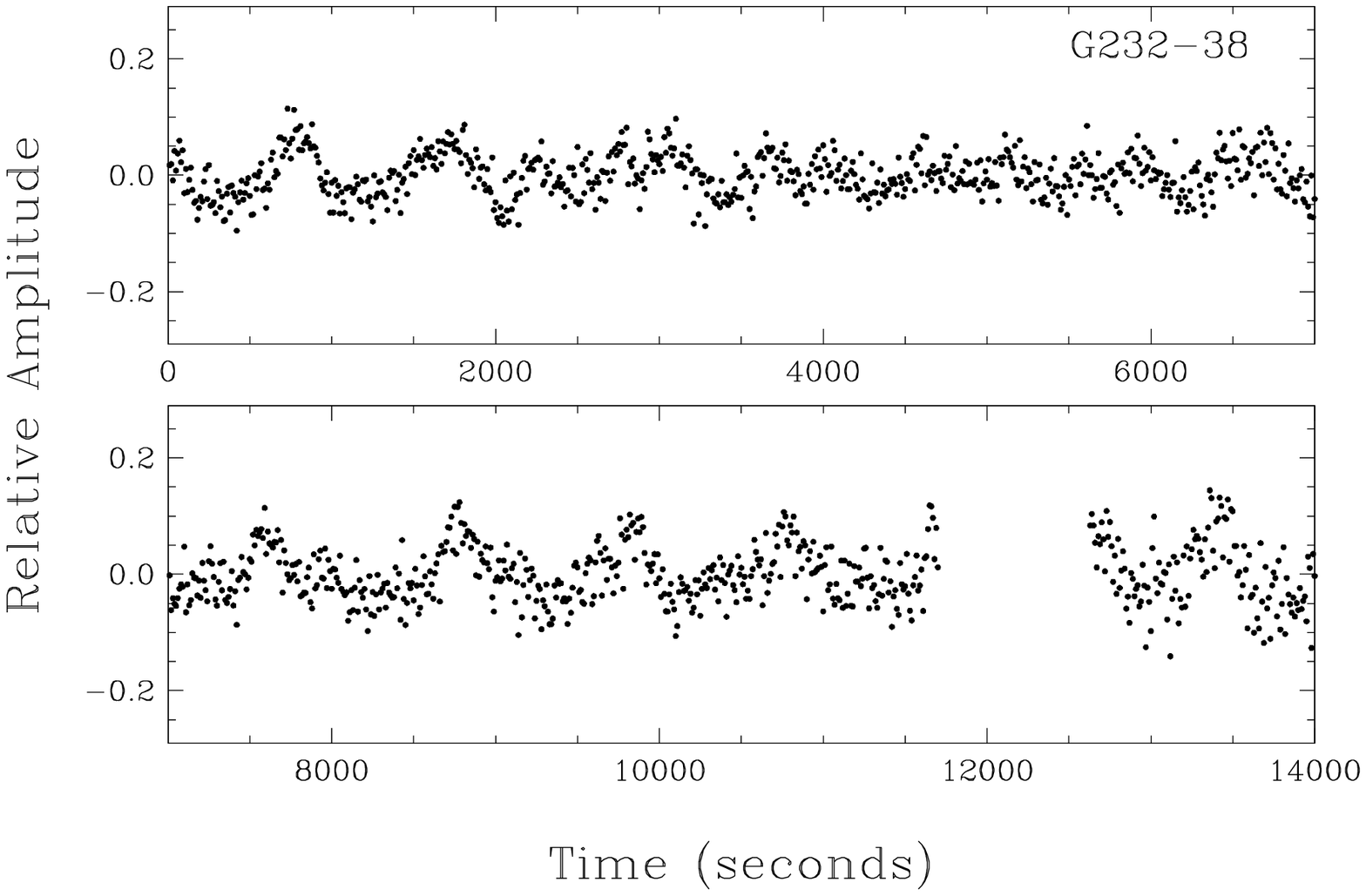] 
{Light curve of G232-38, observed in ``white light'' with LAPOUNE
attached to the Observatoire du mont M\'egantic 1.6 m telescope. Each
point represents a sampling time of 10 s. The light curve is expressed
in terms of residual amplitude relative to the mean brightness of the
star. \label{fg:f2}}

\figcaption[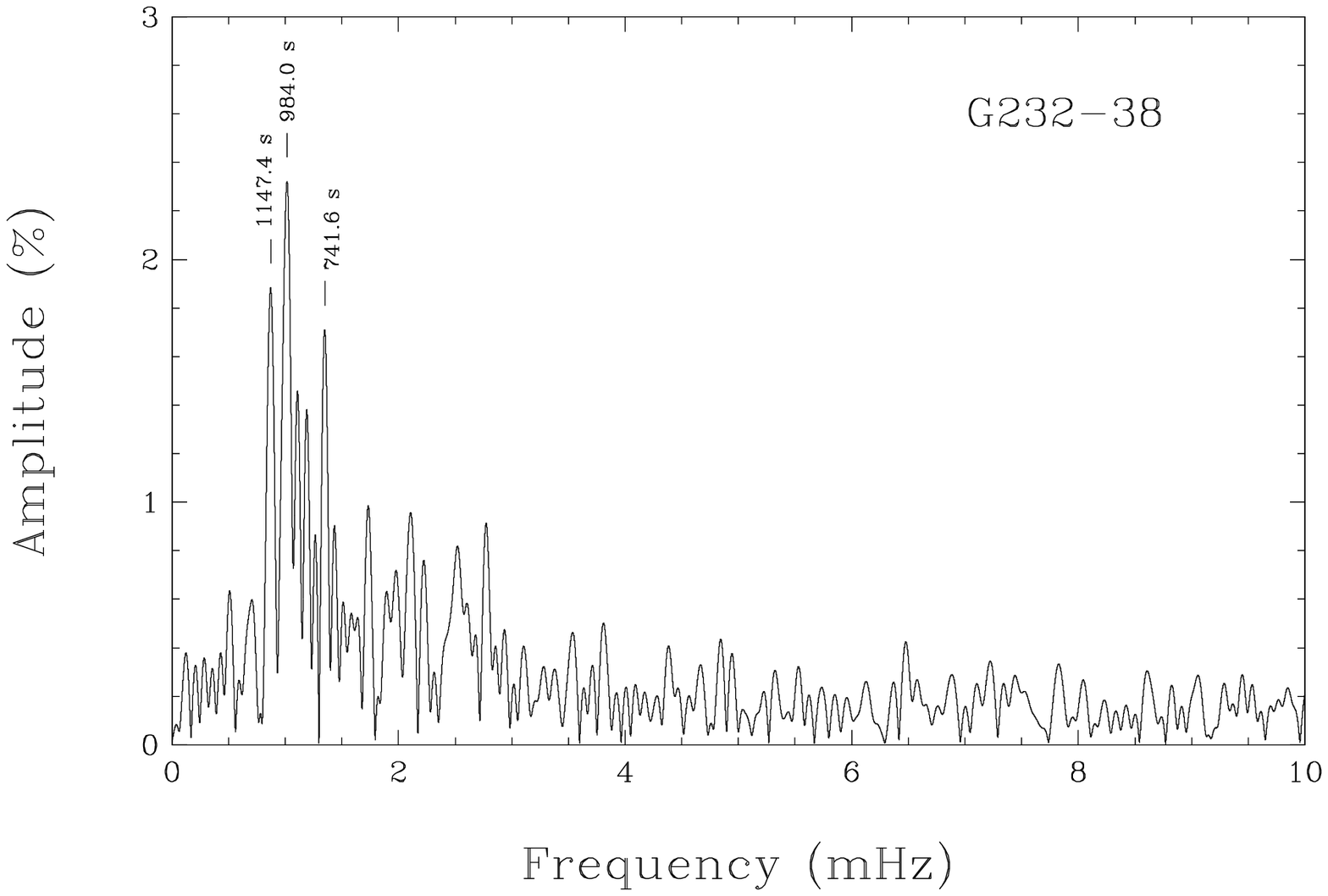] 
{Fourier (amplitude) spectrum of the light curve of G232-38 in the
0-10 mHz bandpass. The spectrum in the region from 10 mHz to the
Nyquist frequency is entirely consistent with noise and is not
shown. The amplitude axis is expressed in terms of the percentage
variations about the mean brightness of the star. \label{fg:f3}}

\figcaption[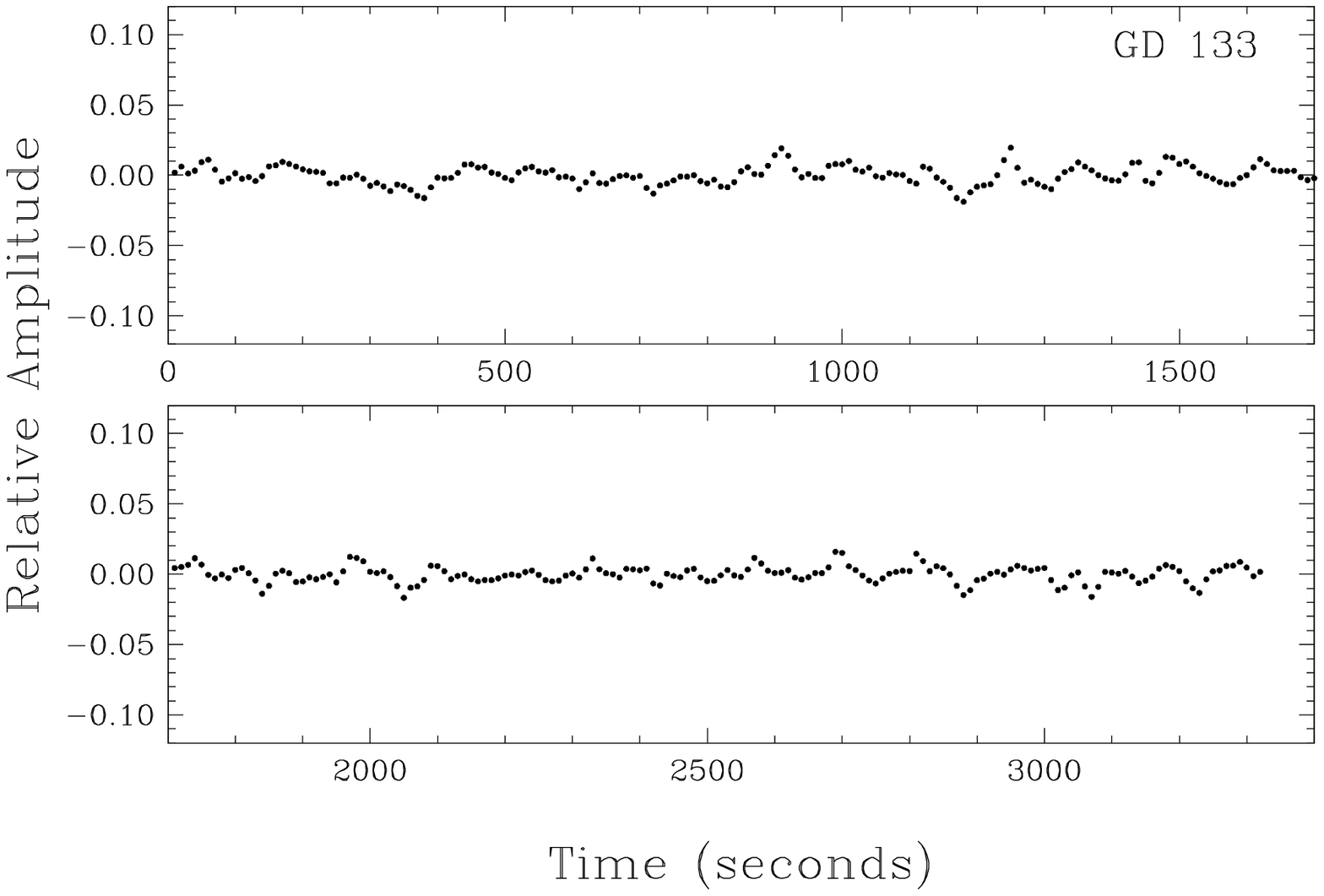]
{Light curve of GD 133, observed in ``white light'' with LAPOUNE
attached to the 61-inch telescope at the Mount Bigelow observatory.
Each point represents a sampling time of 10 s. The light curve is
expressed in terms of residual amplitude relative to the mean
brightness of the star.
\label{fg:f4}}

\figcaption[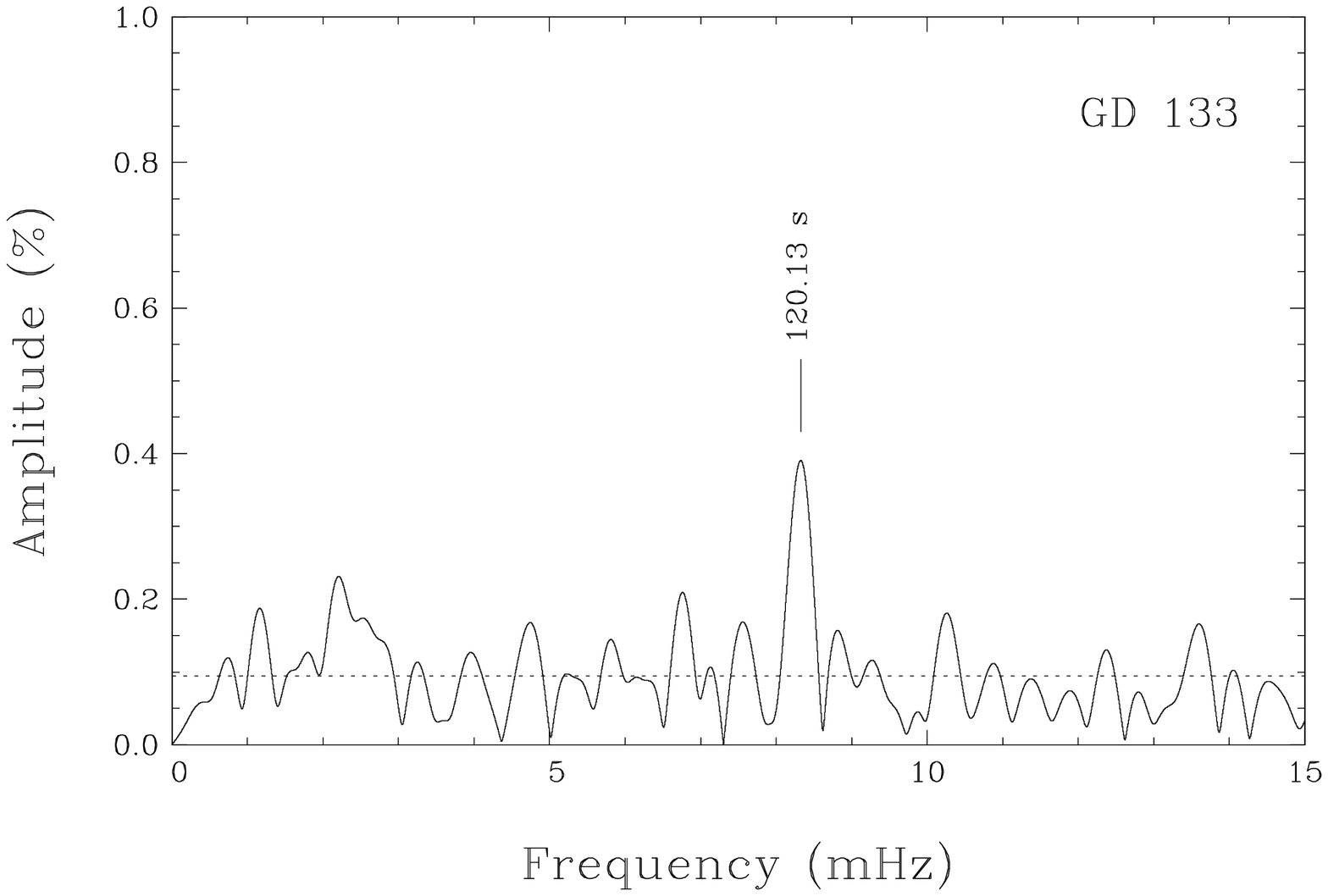]
{Same as Figure \ref{fg:f3} for GD 133 and in the bandpass 0-15
mHz. The dashed line represents the 1$\sigma$ noise level.\label{fg:f5}}

\figcaption[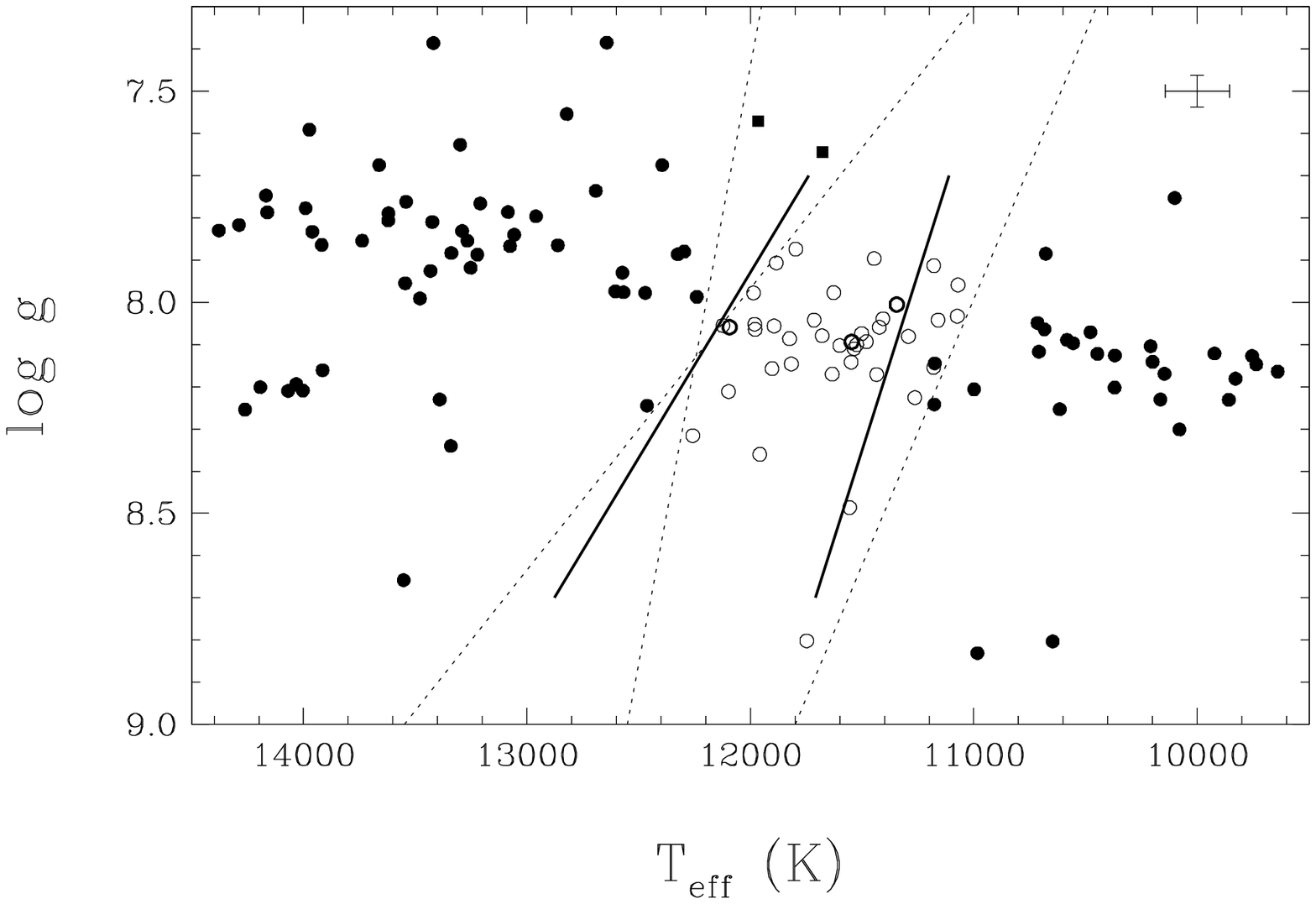] 
{$\Te-\logg$ distribution for DA white dwarfs with high-speed
photometric measurements. The open circles represent the 36 ZZ Ceti
stars from \citet{bergeron04} as well as the three recent discoveries
reported by ({\it bold circles from left to right}) Silvotti et
al.~(2005, in preparation, GD 133), \citet[][PB 520]{silvotti05} and
in this paper (G232-38). The filled circles represent the
photometrically constant DA stars from Table 1 with appropriate
effective temperatures; while the filled squares correspond to
unresolved double degenerate systems. The error bars in the upper
right corner represent the average uncertainties of the spectroscopic
method in the region of the ZZ Ceti stars. The dashed lines represent
the empirical blue and red edges of the instability strip while the
solid lines represent the theoretical boundaries of the instability
strip as computed by
\citet{fontaine03}.\label{fg:f6}}

\figcaption[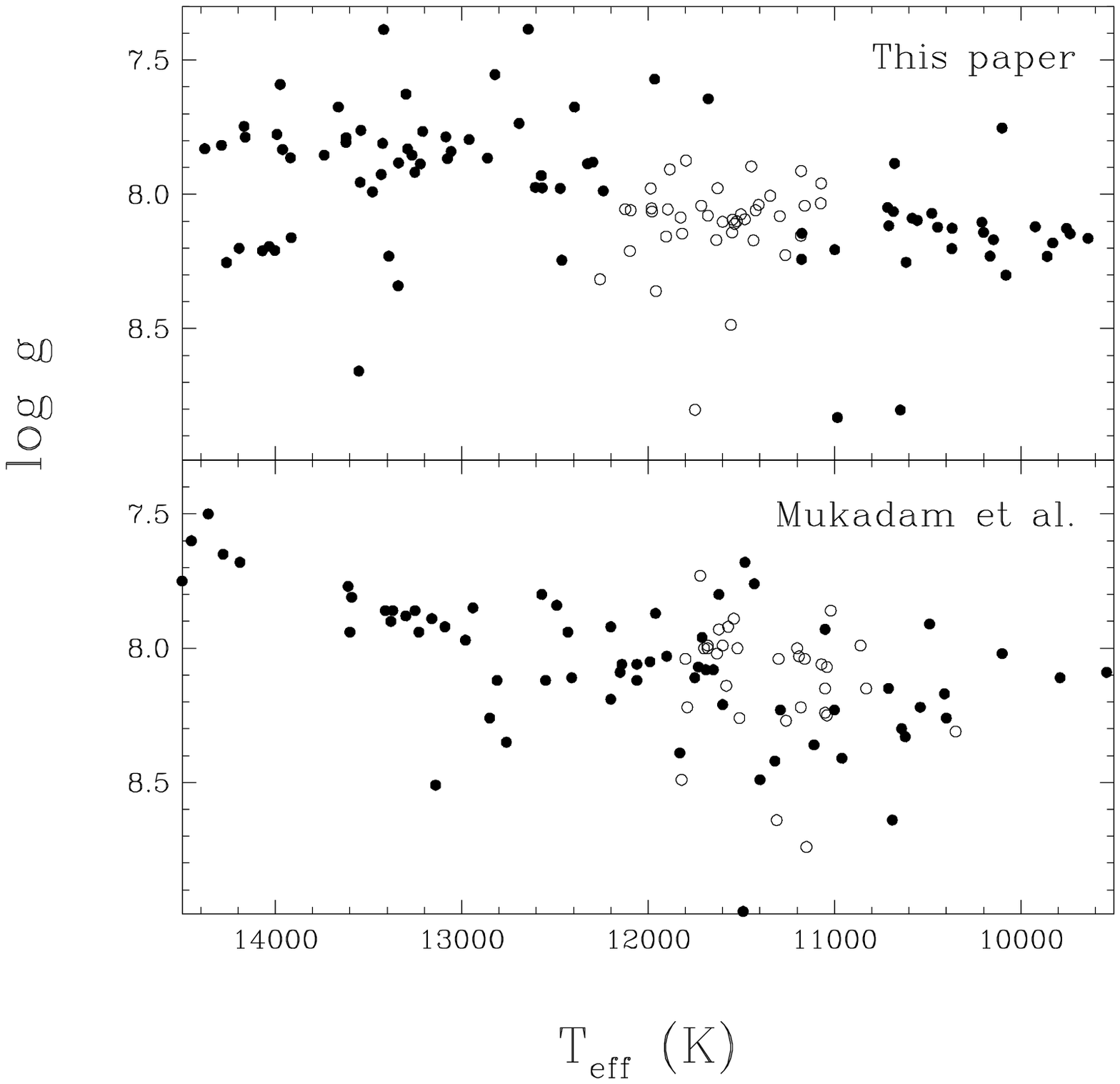] 
{$\Te-\logg$ distribution for DA white dwarfs with high-speed
photometric measurements taken from this paper 
and from the analysis of \citet{mukadam04b}.
The open circles represent ZZ Ceti stars while filled circles
correspond to photometrically constant stars.\label{fg:f7}}

\figcaption[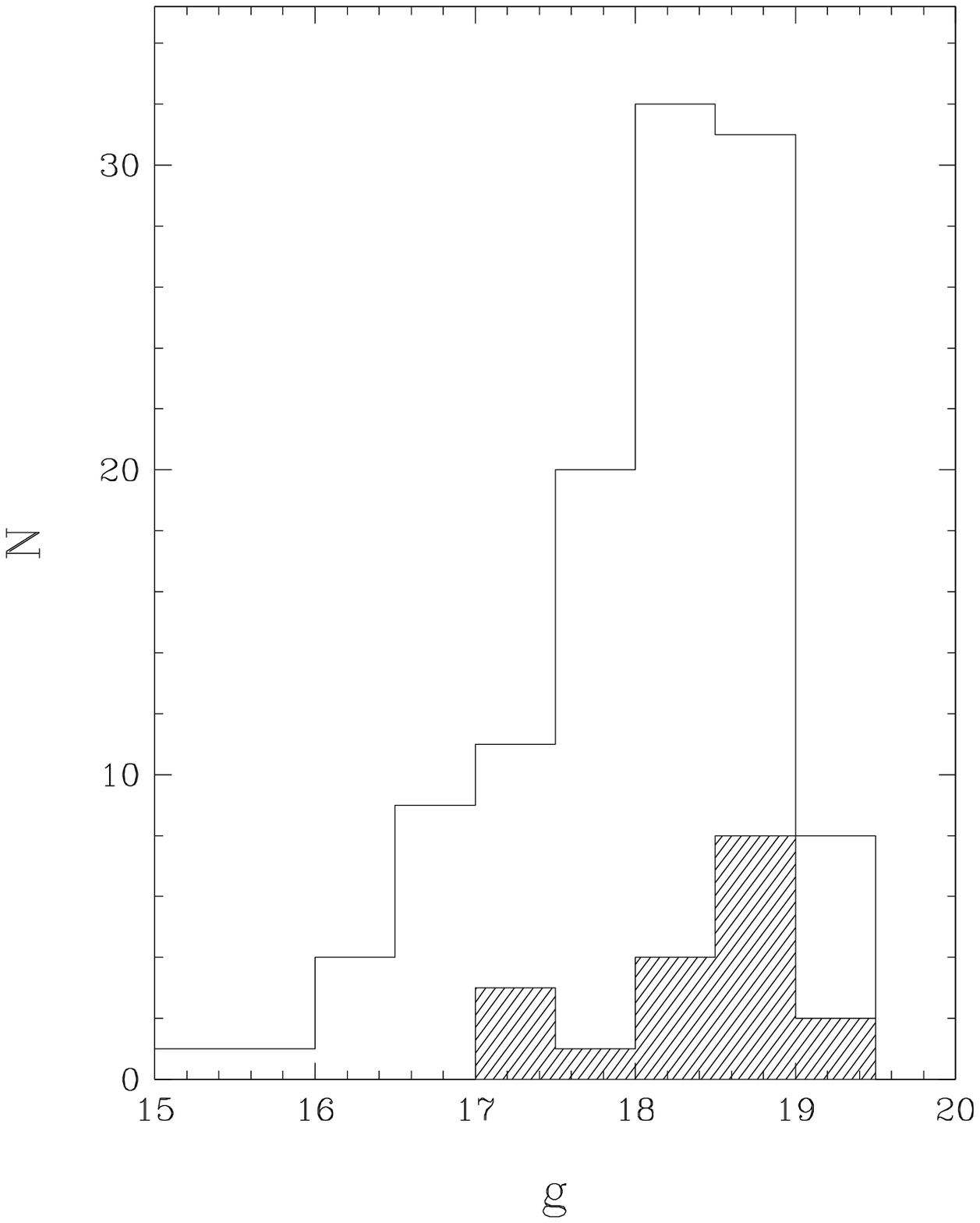] 
{Number distribution of DA stars taken from Tables 1 to 3 of
\citet{mukadam04a} as a function of the $g$ magnitude ({\it solid line
histogram}) compared with the distribution of nonvariables claimed to
lie within the ZZ Ceti instability strip ({\it hatched histogram})
taken from Table 1 of \citet{mukadam04b}. \label{fg:f8}}

\figcaption[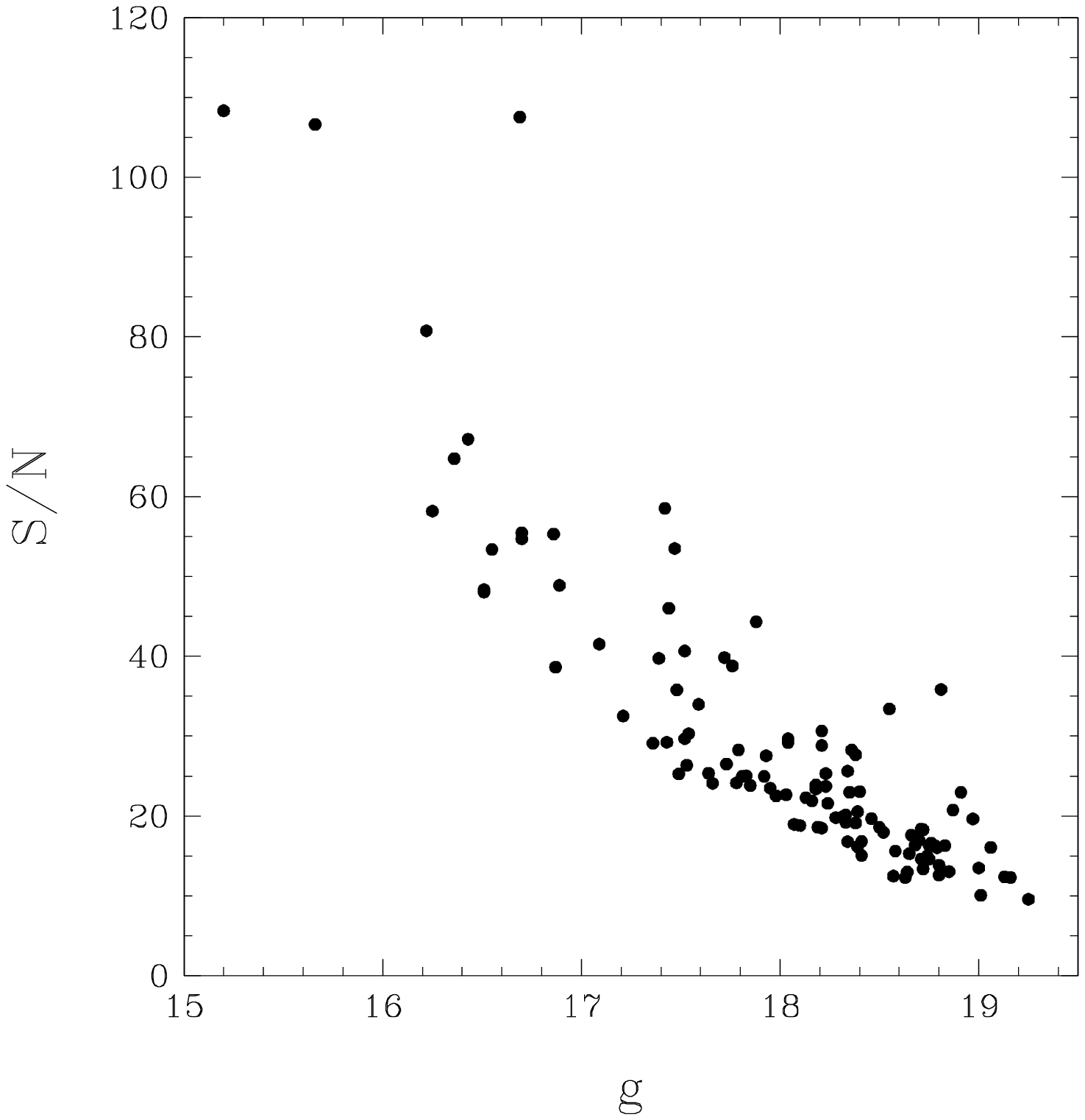] 
{S/N of the spectroscopic observations of the DA
stars from Tables 1 to 3 of \citet{mukadam04a} as a function of
the $g$ magnitude.\label{fg:f9}}

\figcaption[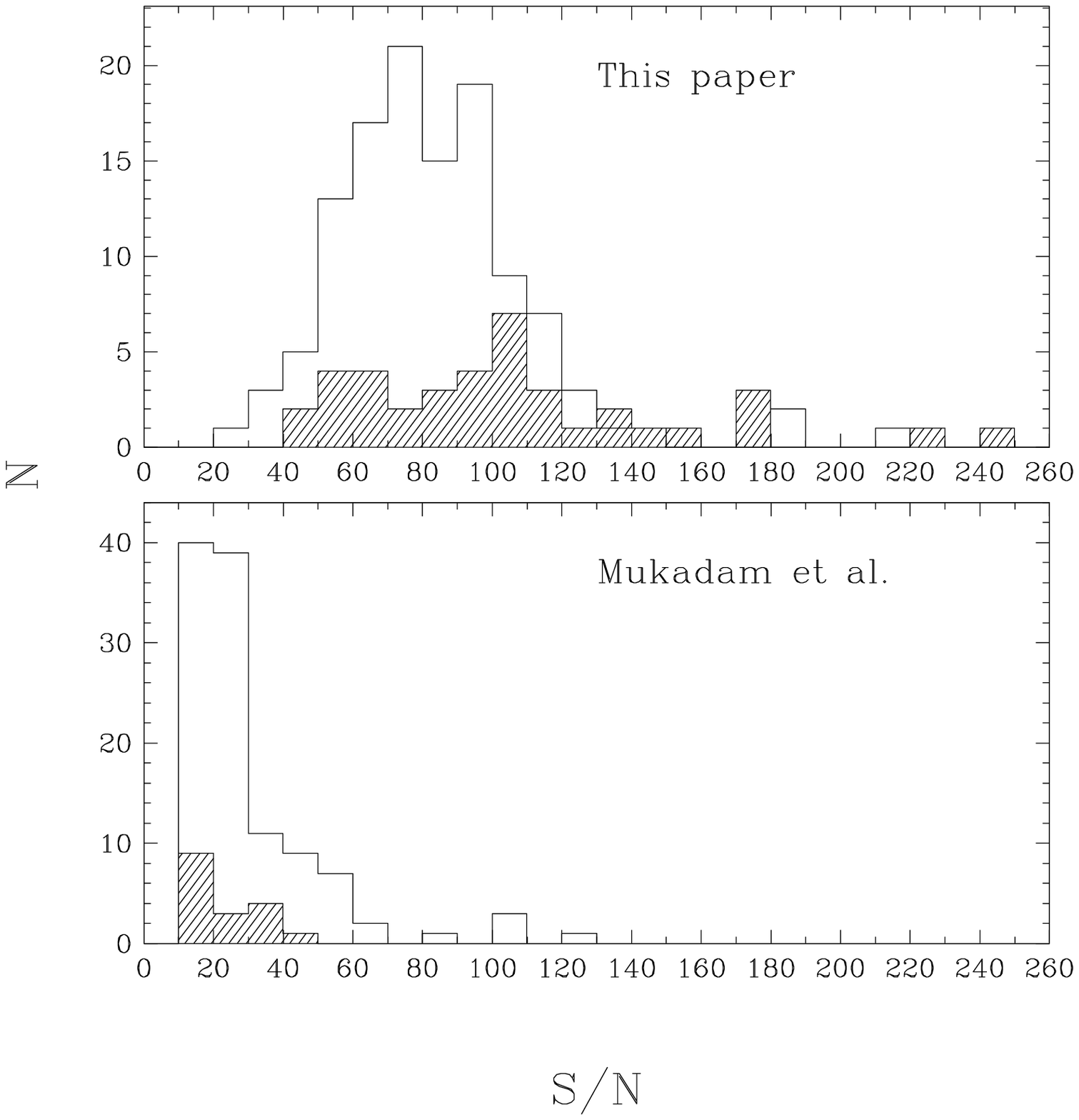] 
{{\it Top:} Distribution of S/N for the 121 spectra of the
photometrically constant stars from Table 1 ({\it solid line
histogram}) and of the 36 ZZ Ceti stars from \citet{bergeron04} and
the 3 additional ZZ Ceti stars from Table 2 ({\it hatched histogram}).  {\it
Bottom:} Same as the top panel but for 113 out of the 118 white dwarf
spectra taken from Tables 1 to 3 of Mukadam et al.~(2004a, {\it solid line
histogram}), and for 17 out of the 18 nonvariables that lie within the
ZZ Ceti instability strip according to Mukadam et al.~(2004b, {\it
hatched histogram}).\label{fg:f10}}

\figcaption[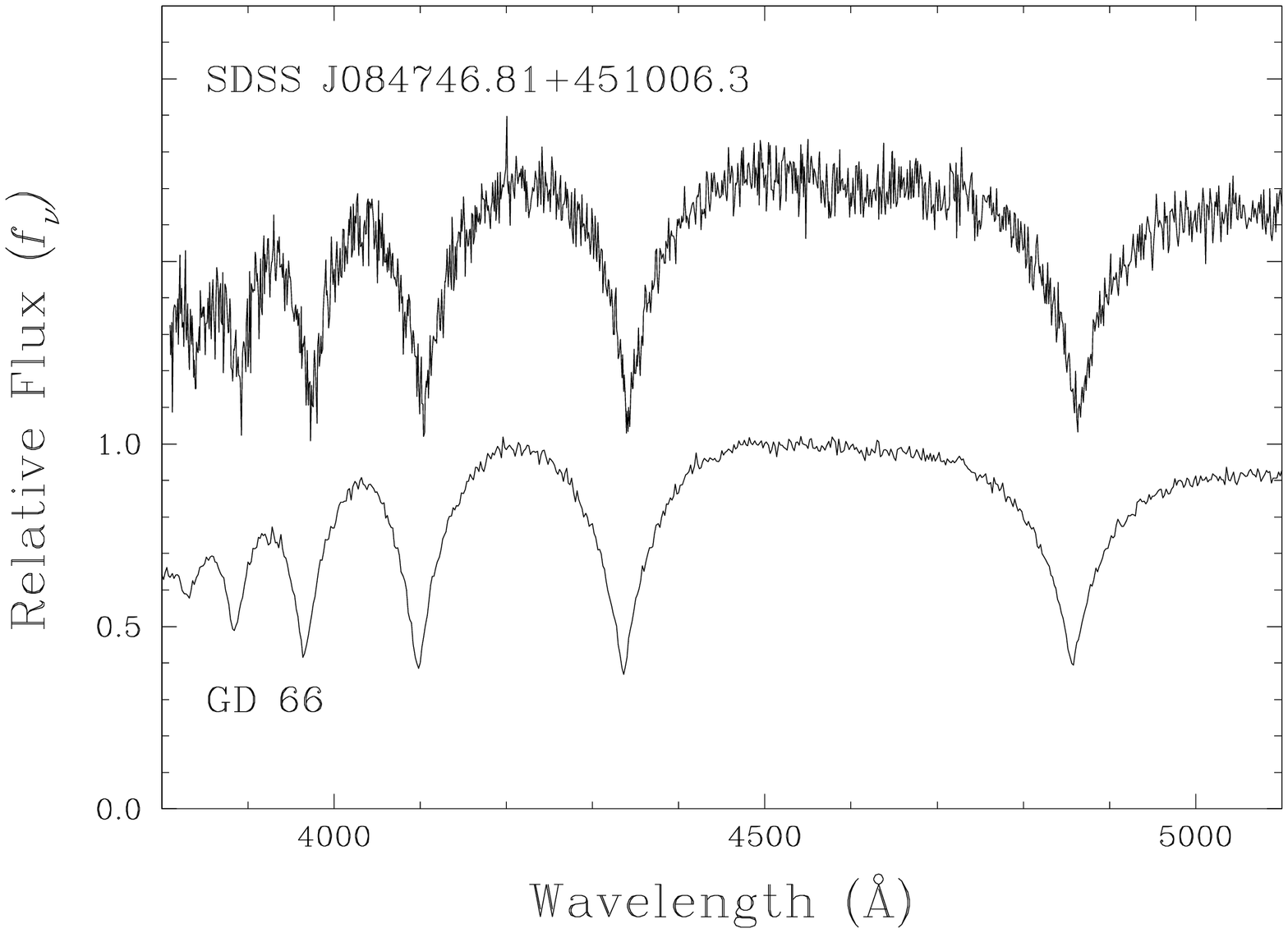] 
{Typical ZZ Ceti spectra from the SDSS ({\it top}) and Bergeron
et al.~(2004, {\it bottom}). Both spectra are flux calibrated,
normalized to unity at 4500~\AA, and offset by a factor of 0.8 for
clarity. The signal-to-noise ratio is
${\rm S/N}=20$ for SDSS J084746.81$+$451006.3 and
${\rm S/N}=80$ for GD 66.\label{fg:f11}}

\figcaption[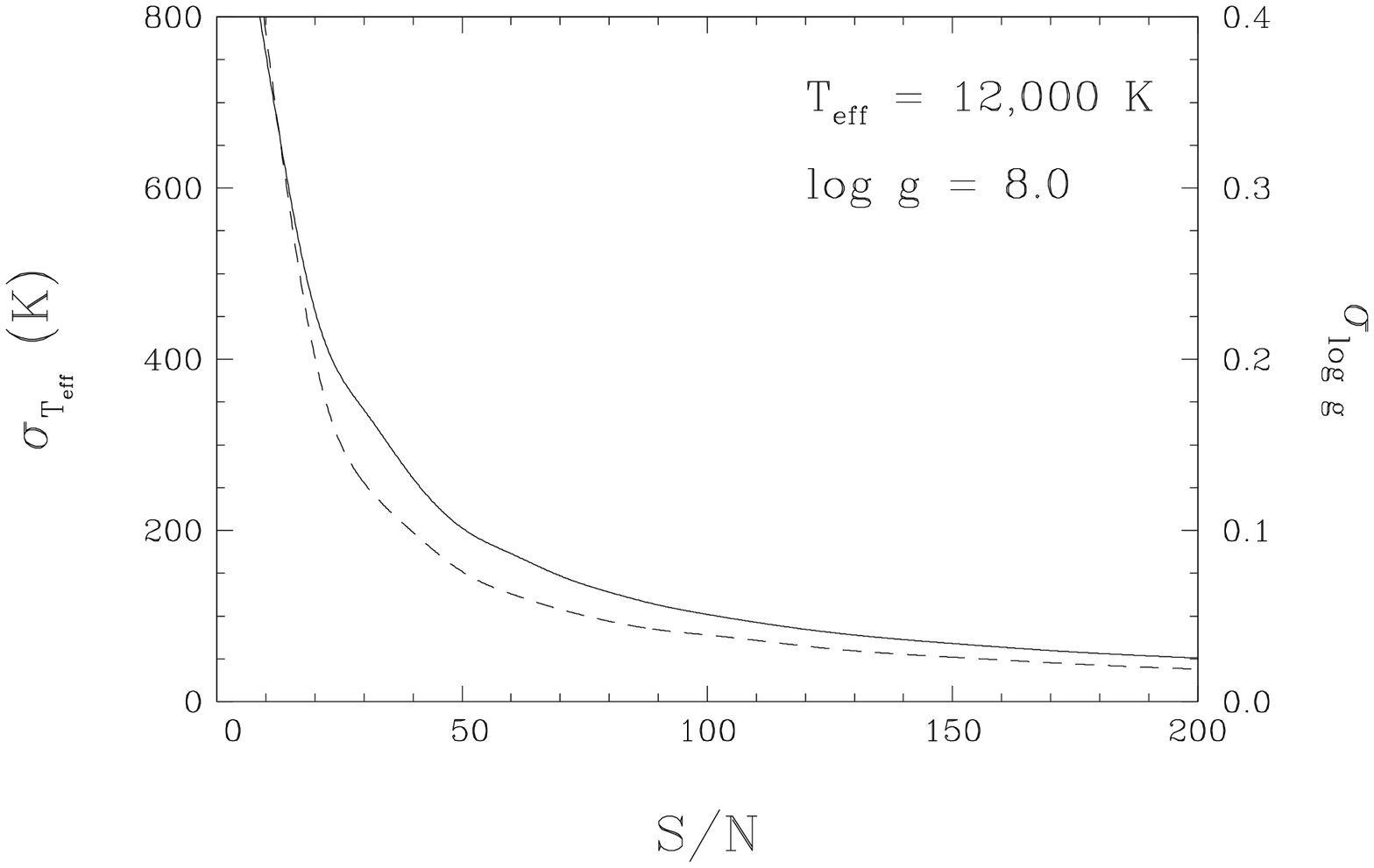]
{A plot of the uncertainty in $\Te$ ({\it solid line}) and $\logg$
({\it dashed line}) as a function of S/N for a simulated (see text) DA
white dwarf with $\Te = 12,000$ K and $\logg = 8.0$.\label{fg:f12}}

\figcaption[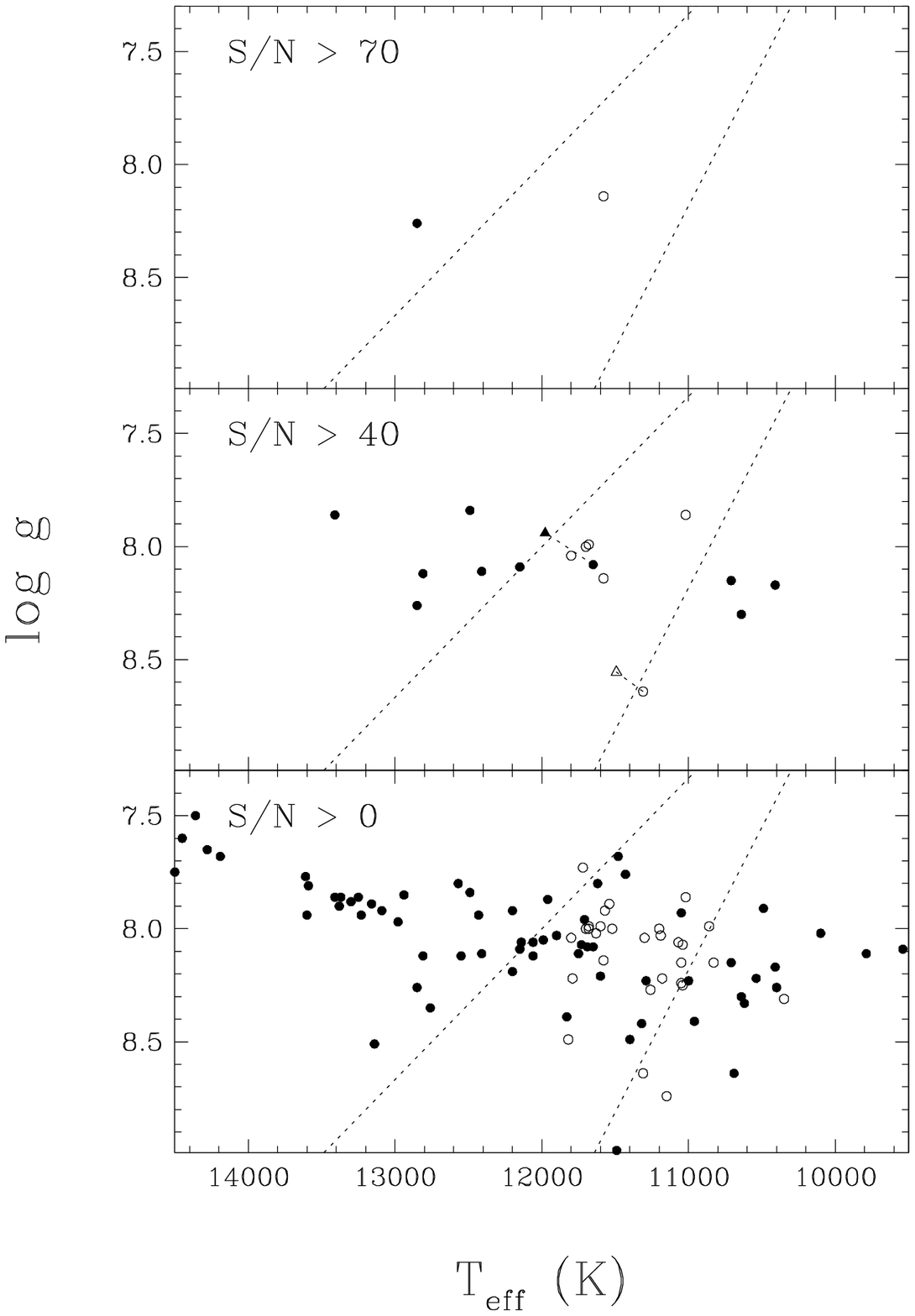] 
{Surface gravity -- effective temperature distribution for the DA
stars taken from \citet{mukadam04a} for different S/N thresholds.  The
dashed lines represent the empirical blue and red edges of the
instability strip defined by \citet{bergeron04}. Open circles
represent ZZ Ceti stars while the filled circles correspond to
nonvariables.  The open and filled triangles represent our
determinations of the atmospheric parameters of WD~1338$-$0023 and
WD~1711+6541 ({\it left and right symbols}, respectively) based on our
own spectroscopic observations and model spectra; dashed lines join
these determinations with those of \citet{mukadam04a}. \label{fg:f13}}

\figcaption[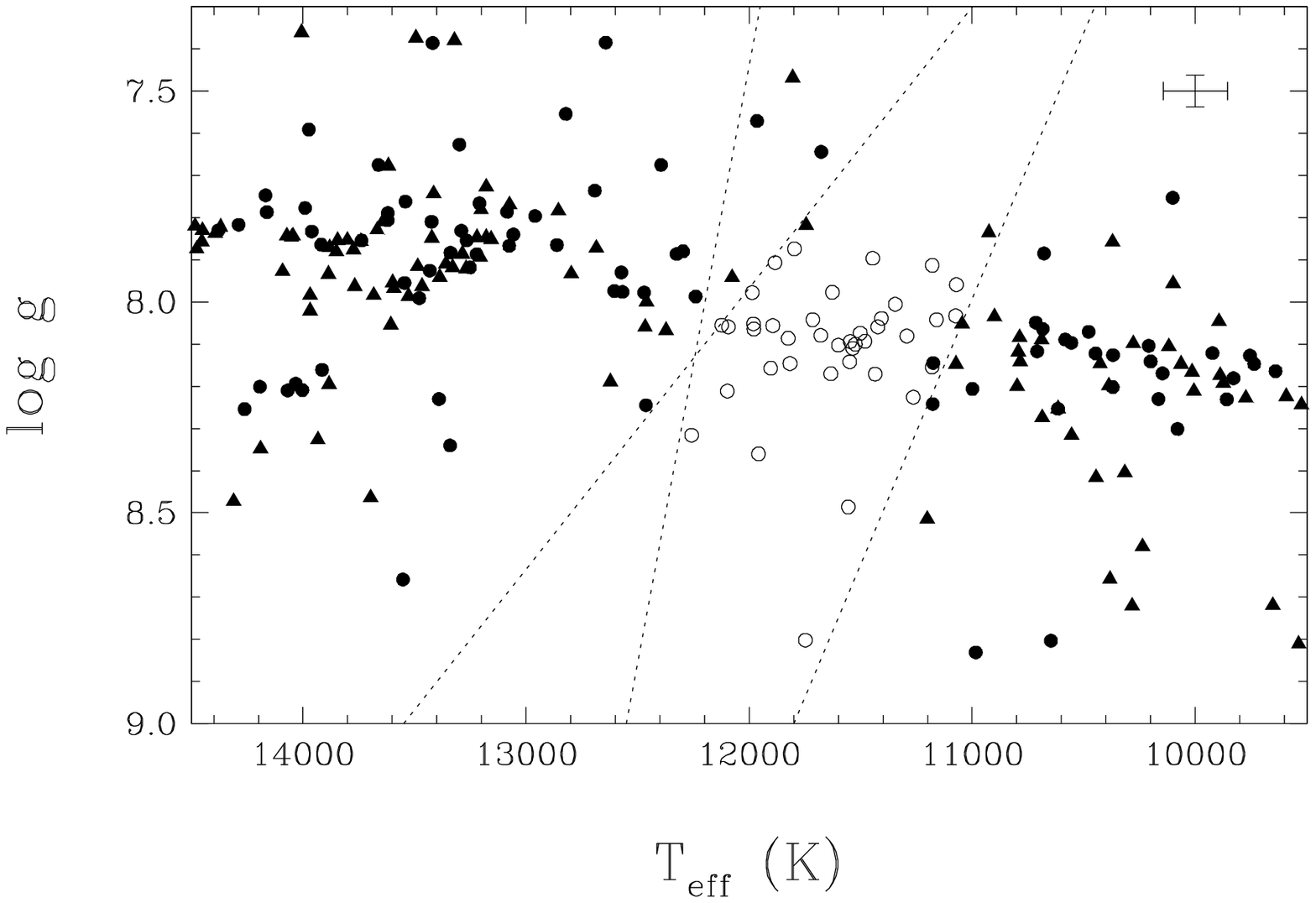] 
{Same as Fig.~\ref{fg:f6} with the addition of all the DA white dwarfs
from our ongoing spectroscopic survey of the catalog of
\citet{mccook99}.  The circles correspond to objects that have been
investigated for photometric variability with the open circles
representing the ZZ Ceti stars, while the filled triangles correspond
to objects that have not been investigated for photometric
variability.\label{fg:f14}}

\clearpage

\begin{figure}[p]
\plotone{f1.eps}
\begin{flushright}
Figure \ref{fg:f1}
\end{flushright}
\end{figure}

\clearpage

\begin{figure}[p]
\plotone{f2.eps}
\begin{flushright}
Figure \ref{fg:f2}
\end{flushright}
\end{figure}

\clearpage

\begin{figure}[p]
\plotone{f3.eps}
\begin{flushright}
Figure \ref{fg:f3}
\end{flushright}
\end{figure}

\clearpage

\begin{figure}[p]
\plotone{f4.eps}
\begin{flushright}
Figure \ref{fg:f4}
\end{flushright}
\end{figure}

\clearpage

\begin{figure}[p]
\plotone{f5.eps}
\begin{flushright}
Figure \ref{fg:f5}
\end{flushright}
\end{figure}

\clearpage

\begin{figure}[p]
\plotone{f6.eps}
\begin{flushright}
Figure \ref{fg:f6}
\end{flushright}
\end{figure}

\clearpage

\begin{figure}[p]
\plotone{f7.eps}
\begin{flushright}
Figure \ref{fg:f7}
\end{flushright}
\end{figure}

\clearpage

\begin{figure}[p]
\plotone{f8.eps}
\begin{flushright}
Figure \ref{fg:f8}
\end{flushright}
\end{figure}

\clearpage

\begin{figure}[p]
\plotone{f9.eps}
\begin{flushright}
Figure \ref{fg:f9}
\end{flushright}
\end{figure}

\clearpage

\begin{figure}[p]
\plotone{f10.eps}
\begin{flushright}
Figure \ref{fg:f10}
\end{flushright}
\end{figure}

\clearpage

\begin{figure}[p]
\plotone{f11.eps}
\begin{flushright}
Figure \ref{fg:f11}
\end{flushright}
\end{figure}

\clearpage

\begin{figure}[p]
\plotone{f12.eps}
\begin{flushright}
Figure \ref{fg:f12}
\end{flushright}
\end{figure}

\clearpage

\begin{figure}[p]
\plotone{f13.eps}
\begin{flushright}
Figure \ref{fg:f13}
\end{flushright}
\end{figure}

\clearpage

\begin{figure}[p]
\plotone{f14.eps}
\begin{flushright}
Figure \ref{fg:f14}
\end{flushright}
\end{figure}

\end{document}